\documentclass[twocolumn]{aastex6}
\usepackage{lineno}

    {\pagebreak[4]\global\pdfpageattr\expandafter{\the\pdfpageattr/Rotate 90}}%
    {\pagebreak[4]\global\pdfpageattr\expandafter{\the\pdfpageattr/Rotate 0}}%

\let\pwiflocal=\iffalse \let\pwifjournal=\iffalse
\usepackage{amsmath,amssymb}
\usepackage{epsfig}    
\usepackage{graphicx}    
\usepackage{lineno}
\usepackage{natbib}
\usepackage{bigints}
\usepackage[outdir=./]{epstopdf}

\usepackage[T1]{fontenc}
\pwifjournal\else
  \usepackage{microtype}
\fi

\pwifjournal\else
  \makeatletter
  \renewcommand\plotone[1]{%
    \centering \leavevmode \setlength{\plot@width}{0.95\linewidth}
    \includegraphics[width={\eps@scaling\plot@width}]{#1}%
  }%
  \makeatother
\fi

\makeatletter

\newcommand\@simpfx{http://simbad.u-strasbg.fr/simbad/sim-id?Ident=}

\newcommand\MakeObj[4][\@empty]{
  \pwifjournal%
    \expandafter\newcommand\csname pkgwobj@c@#2\endcsname[1]{\protect\object[#4]{##1}}%
  \else%
    \expandafter\newcommand\csname pkgwobj@c@#2\endcsname[1]{\href{\@simpfx #3}{##1}}%
  \fi%
  \expandafter\newcommand\csname pkgwobj@f#2\endcsname{#4}%
  \ifx\@empty#1%
    \expandafter\newcommand\csname pkgwobj@s#2\endcsname{#4}%
  \else%
    \expandafter\newcommand\csname pkgwobj@s#2\endcsname{#1}%
  \fi}%

\newcommand\MakeTrunc[2]{
  \expandafter\newcommand\csname pkgwobj@t#1\endcsname{#2}}%

\newcommand{\obj}[1]{%
  \expandafter\ifx\csname pkgwobj@c@#1\endcsname\relax%
    \textbf{[unknown object!]}%
  \else%
    \csname pkgwobj@c@#1\endcsname{\csname pkgwobj@s#1\endcsname}%
  \fi}
\newcommand{\objf}[1]{%
  \expandafter\ifx\csname pkgwobj@c@#1\endcsname\relax%
    \textbf{[unknown object!]}%
  \else%
    \csname pkgwobj@c@#1\endcsname{\csname pkgwobj@f#1\endcsname}%
  \fi}
\newcommand{\objt}[1]{%
  \expandafter\ifx\csname pkgwobj@c@#1\endcsname\relax%
    \textbf{[unknown object!]}%
  \else%
    \csname pkgwobj@c@#1\endcsname{\csname pkgwobj@t#1\endcsname}%
  \fi}

\makeatother

\pwifjournal\else
  \usepackage{etoolbox}
  \makeatletter
  \patchcmd{\NAT@citex}
    {\@citea\NAT@hyper@{%
       \NAT@nmfmt{\NAT@nm}%
       \hyper@natlinkbreak{\NAT@aysep\NAT@spacechar}{\@citeb\@extra@b@citeb}%
       \NAT@date}}
    {\@citea\NAT@nmfmt{\NAT@nm}%
     \NAT@aysep\NAT@spacechar\NAT@hyper@{\NAT@date}}{}{}
  \patchcmd{\NAT@citex}
    {\@citea\NAT@hyper@{%
       \NAT@nmfmt{\NAT@nm}%
       \hyper@natlinkbreak{\NAT@spacechar\NAT@@open\if*#1*\else#1\NAT@spacechar\fi}%
         {\@citeb\@extra@b@citeb}%
       \NAT@date}}
    {\@citea\NAT@nmfmt{\NAT@nm}%
     \NAT@spacechar\NAT@@open\if*#1*\else#1\NAT@spacechar\fi\NAT@hyper@{\NAT@date}}
    {}{}
  \makeatother
\fi

\providecommand{\adsurl}[1]{\href{#1}{ADS}}
\newcommand{\mps}{m\,s$^{-1}$}

\newcommand{\vsini}{$v\sin{i_*}$}

\newcommand{\kepler}{{\it Kepler}}

\newcommand{\um}{$\mu$m}
\newcommand{\fbol}{$F_{\mathrm{bol}}$}

\newcommand\teff{\ensuremath{T_\text{eff}}}

\newcommand\kms{km~s$^{-1}$}

\newcommand{\ktwo}{{\it K2}}

\newcommand\ut{1}
\newcommand\hub{2}
\newcommand\har{4}
\newcommand\nsf{5}

\newcommand\geo{3}
\newcommand\ifa{6}
\newcommand\taurus{7}

\slugcomment{Accepted to AJ}

\shorttitle{Transiting planets in Praesepe}

\shortauthors{Mann et al.}

\begin{document}
 
\title{Zodiacal Exoplanets in Time (ZEIT) IV:\\ seven transiting planets in the Praesepe cluster} 

\author{Andrew W. Mann\footnote{\email{amann@astro.as.utexas.edu}}\altaffilmark{\ut, \hub}, Eric Gaidos\altaffilmark{\geo}, Andrew Vanderburg\altaffilmark{\har,\nsf}, Aaron C. Rizzuto\altaffilmark{\ut}, Megan Ansdell\altaffilmark{\ifa}, Jennifer Vanessa Medina\altaffilmark{\ut,\taurus}, Gregory N. Mace\altaffilmark{\ut}, Adam L. Kraus\altaffilmark{\ut}, Kimberly R. Sokal\altaffilmark{\ut}}

\altaffiltext{\ut}{Department of Astronomy, The University of Texas at Austin, Austin, TX 78712, USA}
\altaffiltext{\hub}{Hubble Fellow}
\altaffiltext{\geo}{Department of Geology and Geophysics, University of Hawaii at Manoa, Honolulu, HI 96822, USA}
\altaffiltext{\har}{Harvard--Smithsonian Center for Astrophysics, Cambridge, Massachusetts 02138, USA}
\altaffiltext{\nsf}{NSF Graduate Research Fellow}
\altaffiltext{\ifa}{Institute for Astronomy, University of Hawaii at Manoa, Honolulu, HI 96822, USA}
\altaffiltext{\taurus}{TAURUS Scholar}

\begin{abstract}
Open clusters and young stellar associations are attractive sites to search for planets and to test theories of planet formation, migration, and evolution. We present our search for, and characterization of, transiting planets in the 800\, Myr old Praesepe (Beehive, M44) Cluster from \ktwo\ light curves. We identify seven planet candidates, six of which we statistically validate to be real planets, the last of which requires more data. For each host star we obtain high-resolution NIR spectra to measure its projected rotational broadening and radial velocity, the latter of which we use to confirm cluster membership. We combine low-resolution spectra with the known cluster distance and metallicity to provide precise temperatures, masses, radii, and luminosities for the host stars. Combining our measurements of rotational broadening, rotation periods, and our derived stellar radii, we show that all planetary orbits are consistent with alignment to their host star's rotation. We fit the \ktwo\ light curves, including priors on stellar density to put constraints on the planetary eccentricities, all of which are consistent with zero. The difference between the number of planets found in Praesepe and Hyades (8 planets, $\simeq800$\,Myr) and a similar dataset for Pleiades (0 planets, $\simeq$125\,Myr) suggests a trend with age, but may be due to incompleteness of current search pipelines for younger, faster-rotating stars. We see increasing evidence that some planets continue to lose atmosphere past 800\,Myr, as now two planets at this age have radii significantly larger than their older counterparts from \kepler.
\end{abstract}


\maketitle

\section{Introduction}\label{sec:intro}
Stellar clusters are unique sites to test theories of planet formation and evolution. Because they consist of chemically homogenous \citep[or nearly homogeneous, ][]{Liu2016} stellar populations, planets in clusters facilitate statistical studies of exoplanet properties (e.g., occurrence, period, size) while controlling for changes due to host star metallicity \citep[e.g.,][]{2012Natur.486..375B,Neves:2013yq,Mann2013b}. Common and well-measured age, metallicity, and distance help yield more precise determinations of stellar parameters for cluster members than is generally possible for field stars, often providing a proportional improvement in planet parameters. The comparatively well constrained ages available for clusters compared to field stars \citep[e.g.,][]{2014ApJ...780..117S, 2014ApJ...782...29C,2015ApJ...813..108D} also facilitate studies of planetary evolution. To this end, young ($<1$\,Gyr) clusters are particularly useful because planetary systems undergo the most change in the first few hundred megayears \citep[e.g.,][]{Adams2006, 2009ApJ...696L..98R}. The {\it Kepler} mission \citep{Borucki2010} has found thousands of (candidate) planets \citep[e.g.,][]{2015ApJS..217...31M}. However, target stars in the {\it Kepler}-prime mission are generally older than 1\,Gyr, and most have poorly constrained ages \citep{Batalha:2010fk, 2013MNRAS.436.1883W, 2015MNRAS.452.2127S}, making them less useful for evolutionary studies. Planets in young stellar clusters could fill this gap. 

Because of their scientific value, open clusters have long been targeted for exoplanet searches \citep[e.g.,][]{Cochran2002,2002AJ....123.3460M,2007MNRAS.375...29A,SandersGaudi2011}. Despite numerous surveys, only a handful of planets in open clusters were discovered prior to \kepler\ \citep[e.g.,][]{2007A&A...472..657L,2012ApJ...756L..33Q}, and none were significantly smaller or less massive than Jupiter. These early searches were generally only sensitive only to Jovian planets on close orbits, which are intrinsically rare \citep[e.g.,][]{Johnson:2010lr,Gaidos2013b,Fressin:2013qy}. Most nearby clusters ($<200$\,pc) are younger than 1\,Gyr, and thus their members are noisier (in terms of radial velocity (RV) and photometric variation) than their older counterparts \citep{2004AJ....127.3579P,2010ApJ...710..432R,Crockett2012}, which complicates the detection and characterization of any planetary signal. Because of brightness limitations, studies of more distant, older clusters were limited to brighter F-, G-, and early-K-type members, which removes $>70$\% of the potential targets and makes detecting even a few planets unlikely. Though recent improvements in sensitivity may enable RV surveys to detect down to Neptune-mass objects on close orbits in the coming years \citep{2016A&A...588A.118M,Quinn2014}. 

The {\it Kepler} spacecraft can detect the much more common Earth-to-Neptune-size planets \citep{Jenkins:2010qy}, which has enabled the discovery of two planets smaller than Neptune in the open cluster NGC6811 \citep{Meibom2013}. However, NGC6811 is $\sim 1$~Gyr old, and resides at a distance of 1100\,pc, which provides limited temporal information and makes follow-up difficult. The {\it Kepler}-prime field contains no open clusters clusters that are significantly younger or closer than NGC6811 within which to search for planets. 

The repurposed \kepler\ mission \ktwo, \citep{Howell2014} provides a unique opportunity to revisit open clusters for planet searches. So far \ktwo\ has observed Praesepe, Hyades, M35, and Pleiades, the young star-forming regions Upper Scorpius and $\rho$ Ophucus, and older clusters M67 and Ruprecht 147. {\it K2} observations of Taurus-Auriga and additional visits to Hyades, Upper Scorpius, and Praesepe are planned for future campaigns. They span ages of $\simeq2$ to 800\,Myr, supplying an unparalleled dataset to explore planetary (and stellar) evolution. These groups are all sufficiently close ($<200$\,pc) to search for planets around the more numerous M-dwarf members. Furthermore, because \ktwo\ is sensitive to super-Earth and Neptune-size planets \citep{Dressing2013,2013ApJ...770...69P,Gaidos2016b} around most target stars \citep[e.g.][]{2016ApJS..222...14V,Crossfield:2016aa} the expected planet yield is much higher than earlier surveys.

To take full advantage of the \ktwo\ data set, we launched the Zodiacal Exoplanets in Time (ZEIT) survey. Our aims are to identify, characterize, and explore the statistical properties of planets in nearby young clusters and star-forming regions utilizing both \ktwo\ light curves and a suite of ground-based instruments for follow-up. Our long-term goal is to gain a better understanding of the evolution of planets from infancy ($<$10 Myr) to maturity ($>$1Gyr), including changes in their physical properties, dynamics, and atmospheres. Thus far, we have identified two planets, one in the $\simeq$800\, Myr old Hyades cluster \citep{Mann2016a}, and one in the $\simeq$11\, Myr old Upper Scorpius OB association \citep{Mann2016b} \citep[see also,][who independently discovered both systems]{David2016,David2016b}. Our search of the $\simeq125$\, Myr old Pleiades cluster data turned up only a single planet, which is more likely to be a young field star with Pleiades-like kinematics than a true cluster member \citep{Gaidos:2017aa}. These planets represent important age benchmarks,and can be used to improve our understanding of planetary evolution, but the inclusion of only two planets is insufficient for statistical work. 

Here we present our search for, and characterization of, planets in the Praesepe cluster (also known as the Beehive cluster or M44). In total, we found seven planet candidates, which we follow-up with ground-based spectroscopy and adaptive optics imaging of the host stars (Section~\ref{sec:obs}). We combine these data with literature photometry and astrometry to constrain the properties (mass, radius, etc.) of each host star and confirm their membership to Praesepe (Section~\ref{sec:params}). In Section~\ref{sec:transitfit} we describe our fit of the transit light curves in order to constrain planetary properties, including eccentricity. We use publicly available software to assess the false-positive probability of each system in Section~\ref{sec:fpp}, with which we confirm the planetary nature of six out of seven planet candidates. We conclude in Section~\ref{sec:discussion} with a discussion of the dynamical state, frequency, and size of the Praesepe and Hyades planets when compared to significantly older systems.

\section{Observations and Data Reduction}\label{sec:obs} 

\subsection{\ktwo\ Observations and Transit Identification}\label{sec:k2}
From 2015 April 27 to 2015 October 31 (Campaign 5), \ktwo\ observed $\sim$900 known members of the Praesepe cluster. Owing to the loss of two reaction wheels, the \kepler\ spacecraft drifts on $<$day timescales \citep{2016PASP..128g5002V}. To correct the pointing, \kepler's thrusters fire every $\sim$6\,hours. However, during the drift and subsequent thruster fire, stellar images will drift with respect to the detector. Combined with variations in the pixel sensitivity, this drift generates changes in total measured flux from a given star as a function of centroid position. 

Multiple methods have been implemented to mitigate or remove noise from {\it K2} drift. We utilized both `K2SFF' \citep{Vanderburg2014} and 'K2SC' \citep{2016MNRAS.459.2408A} corrected light curves for all targets. K2SFF curves were corrected for noise due to telescope drift by correlating flux measurements with the spacecraft's pointing. K2SC curves were derived using Gaussian Process regression to model changes that depend on the target's position (flat field variability) and time (stellar variability) simultaneously. We also extracted our own light curves, following the method of \citet{Vanderburg2014}, but including a simultaneous fit for stellar variability with a lower (1\,day) cutoff on the stellar rotation period than allowed by \citet{Vanderburg2014}. Some Praesepe-age M-dwarfs will have rotation periods shorter than this \citep{2014ApJ...795..161D}, but {\it K2} long-cadence observations yield too few data points for significantly shorter period cutoffs, and fitting out variation on $\sim$hour timescales runs the risk of removing or altering long-duration transits. 

We downloaded (or extracted from the pixel data) light curves for all Praesepe members given in \citet{Kraus:2007yq} observed by \ktwo\ from the Barbara A. Mikulski Archive for Space Telescopes (MAST). We ran a box least-squares \citep{Kovacs2002} search for transits on each light curve after correcting for stellar variability. Additional details on our search method are given in \citet{Gaidos:2017aa}. Eclipsing binaries were identified visually for separate analysis (A. L. Kraus et al., in preparation). Other artifacts from poorly corrected stellar variability, flares, or red noise are flagged by identifying changes in the transit shape and depth with time, comparing the planet candidate's orbital period to the stellar rotation period, and examining the transit by eye (though no candidates were rejected through visual examination alone). We used our own curves to verify that candidate signals were not artifacts of the reduction process. If a planet was identified in K2SFF or K2SC light curves and not in our curves we re-extracted the relevant light curve, manually locking the stellar rotation correction to the value derived from a Lomb--Scargle periodogram, and removing outliers manually. It is infeasible to repeat this process for all target stars, but it is simple to do on the few with potential signals. In this way all candidate signals were eventually identified with our own light curve. 

In total, seven planet candidates survived our vetting process. Four of these planets have been previously identified by earlier analyses of the \ktwo\ data \citep{2016MNRAS.tmp.1056L,Barros:2016aa,2016MNRAS.461.3399P}, two of which were recognized as orbiting Praesepe members \citep{2016MNRAS.tmp.1056L}, and one of which (K2-95) was characterized in detailed and confirmed to be planetary by \citet{Obermeier:2016aa}. 

For each of the candidates, we extracted a new light curve after the transit was identified. Re-extraction was done because corrections for {\it K2} pointing drift and stellar variability may incorrectly fit out or otherwise negatively affect the transit \citep{Grunblatt:2016aa}. Once the transit is identified, we can include this in the fit to eliminate or mitigate systematic errors introduced this way. Following \citet{Becker2015} and \citet{Mann2016a}, we simultaneously fit for low frequency variations from stellar activity, \kepler\ flat field (drift), and the transits of each system using a least-squares minimization. Both stellar variability and the effect of errors in detector response were modeled as splines as a function of time and centroid position with breakpoints every 0.2~days and 0.4\arcsec, respectively. We used the re-extracted and flattened light curves for measuring transit properties (Section~\ref{sec:transitfit}), but used light curves with only basic processing (flat field/drift correction) for measuring stellar rotation periods.

\subsection{Optical Spectra from SNIFS}
On 2016 January 17 (UT), we obtained an optical spectrum of each target with the SuperNova Integral Field Spectrograph \citep[SNIFS,][]{Aldering2002,Lantz2004} on the University of Hawai'i 2.2m telescope on Maunakea. SNIFS covers 3200--9700\,\AA\ simultaneously with a resolution of $R\simeq$700 and R$\simeq$1000 in the blue (3200--5200\,\AA) and red (5100-9700\,\AA) channels, respectively. Exposure times varied from 60 to 1800s, providing a typical S/N$=$90 per resolving element near 6500\AA. ThAr arcs were taken before or after each observation, which helps to extract the spectrum and improve the wavelength solution. Bias, flat, dark correction and cosmic-ray rejection, construction of the data cubes, and extraction of the one-dimensional spectrum are described in detail in \citet{Aldering2002}. We observed spectrophotometric standards throughout the night, which were used in conjunction with a model of the atmospheric absorption above Maunakea to telluric correct and flux calibrate the spectrum. More details on our observing and reduction methods can be found in \citet{Mann2015b}.

\subsection{Near-infrared Spectra with SpeX}
During the nights of 2016 January 29, February 21, or March 5 (UT), we obtained a spectrum of each target with the near-infrared (NIR) spectrograph SpeX, mounted on the Infrared Telescope Facility on Maunakea. Observations were taken in cross-dispersed (SXD) mode with the $0.3\arcsec$ slit, yielding a resolution of $\simeq$2000 with complete coverage from 0.7--2.5\um. Each target was placed on two positions (A and B) on opposite ends of the slit. After each integration, the object was nodded following an ABBA pattern. Image differences (A$-$B) were used to subtract emission from the atmosphere. Integration times varied based on the brightness of the target, but all were capped at 120s per exposure to mitigate atmospheric variations. For fainter targets, more ABBA sequences were taken until the desired S/N was reached. For all targets we obtained an S/N per resolving element of $>60$ in the center of the $H$ and $K$ bands. Internal flat and arc lamps and A0V standards were observed for each target at a similar airmass and sky position as the target. Individual spectra were reduced, extracted, and stacked using the \textit{SpeXTool} package \citep{Cushing2004}. Telluric correction and flux calibration were applied using the A0V standard and the \textit{xtellcor} package \citep{Vacca2003}.

\subsection{High-resolution Spectra with IGRINS}\label{sec:IGRINS}
We observed each of the seven planet hosts during the nights of 2016 February 24 or April 21 (UT) with the Immersion Grating Infrared Spectrometer \citep[IGRINS,][]{Park2014,Mace2016} on the 2.7m Harlan J. Smith telescope located at McDonald Observatory. IGRINS uses a silicon immersion grating \citep{2010SPIE.7735E..1MY} to achieve high resolving power ($R\simeq$45,000) and simultaneous coverage of both $H$ and $K$ bands (1.48-2.48\,\um) on two separate Hawaii-2RG detectors. IGRINS is stable enough to achieve RV precision of $\lesssim$40\,m\,s$^{-1}$ by using telluric lines for wavelength calibration.

Due to a higher false-positive probability (see Section~\ref{sec:fpp}) and a transit shape consistent with a grazing eclipsing binary, we obtained three additional epochs of one target (EPIC 211901114). These were taken on 2016 October 10, 11, and 12 (UT) with IGRINS on the Discovery Channel Telescope. 

All observations were taken following commonly used strategies for point-source observations with IGRINS. To briefly summarize, each target was placed at two positions along the slit (A and B), taking an exposure at each position in an ABBA pattern as with the SpeX observations. Exposure times varied based on the target's $K_S$ magnitude, but were capped at 600s to avoid saturation of sky emission lines. For the faintest targets, additional ABBA sequences were taken until the required S/N was achieved. To help remove telluric lines, A0V standards were observed following the same pattern. Enough A0V standards were taken to ensure there was at least one standard taken within 0.1 airmasses and 1 hour (of time) of every target.

The IGRINS spectra were reduced using version 2.1 of the publicly available IGRINS pipeline package\footnote{https://github.com/igrins/plp} \citep{IGRINS_plp}, which includes flat-fielding, background removal, order extraction, distortion correction, wavelength calibration, and basic telluric correction using the A0V standards and an A star atmospheric model. Spectra without telluric corrections applied were preserved and used to improve the wavelength solution and provide a zero-point for the RVs. 

\subsection{Adaptive Optics Imaging and Aperture Masking} 
During the nights of 2016 March 19, and March 22, (UT), we observed four of the seven planet hosts (K2-100, K2-101, K2-102, and K2-103) using natural guide star (NGS) adaptive optics (AO) imaging \citep{2000PASP..112..315W} and non-redundant aperture masking (NRM). The three other targets (K2-95, K2-104, and EPIC 211901114) are too faint for NGS, and the Keck 2 laser was not operational during these two nights. 

All observations were taken with the facility imager, NIRC2, on Keck II atop Maunakea. Vertical angle mode was used for both imaging and NRM observations, always utilizing the smallest pixel scale ($9.952 \pm 0.002$ mas/pix). Imaging was taken with the $K'$ or $K_c$ (for K2-100) filter and masking with the nine-hole mask. After AO loops closed on each target, we took four to eight images, adjusting coadds and integration time based on the brightness of the target. For NRM, we took six interferograms, each with an integration of time of 20\,s and a single coadd.

Data reduction and analysis was done following \citet{Kraus2016a}. To summarize, each frame was linearized and corrected for distortion using the NIRC2 solution from \citet{Yelda2010}, then dark and flat corrected using calibration data taken the same night. We interpolated over ``dead'' and ``hot'' pixels, which were identified from superflats and superdarks built from data spanning 2006 to 2013. Pixels with flux levels $>10\sigma$ above the median of the eight adjacent pixels (cosmic rays) were replaced with the median. We searched for faint and wide companions in the AO imaged by first subtracting an azimuthal median PSF model. Close-in companions were identified by first constructing and subtracting a best-fit PSF of another (single-star) taken on one of the two observing nights. All images of a given target were stacked, and searched for companions using 40\,mas radius apertures centered on each pixel. Detection limits were determined from the standard deviation of the flux among all apertures.

Reduction of masking observations follows the appendix of \citet{Kraus2008}. To remove systematics, the observation of each target was paired with a calibration observation of another nearby member of Praesepe or known single-star calibrator taken from \citet{Hartkopf:2001}. Binary system profiles were then fit to the closure phase to produce detection limits. More details on the reduction of masking data can be found in \citet{Kraus2008} and \citet{Kraus2016a}.

\begin{figure}
 	\centering
 	\includegraphics[width=0.46\textwidth]{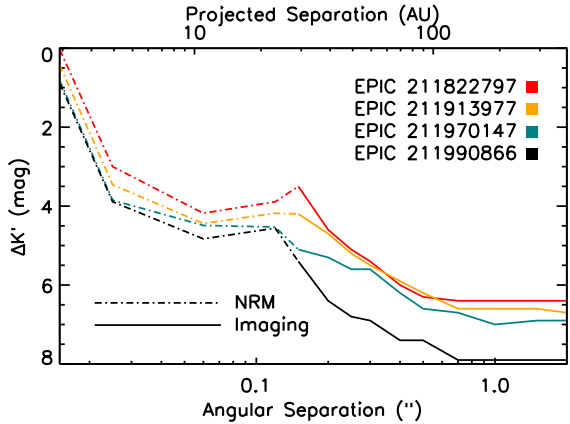} 
 	\caption{Detection limits (5$\sigma$)  as a function of separation for the four targets with AO imaging and NRM interferometry. The top axis shows the separation in AU assuming a distance of 182\,pc. The region probed by non-redundant aperture masking is marked with a dotted--dashed line, while the region probed by imaging is solid. The contrast limits are approximately flat for separations $>3\arcsec$, but due to finite chip size are incomplete as a function of azimuthal angle. }
 	\label{fig:ao}
 \end{figure}

Detection limits (in terms of contrast ratio) as a function of separation constructed from the combination of masking and imaging for the four targets observed are shown in Figure~\ref{fig:ao}. Owing to the edges of the detector the azimuthal coverage is not complete past $\simeq$3\arcsec\ depending on where the object was placed on the detector. Only one target had a significant detection; the images of K2-100 show a faint ($\Delta K' = 5.830\pm0.010$) companion at a separation of $1017.7 \pm 1.6$\,mas and a position angle of $98.623 \pm 0.090$\,degrees. The companion is close enough to land in the selected \ktwo\ aperture (pixel size $=3.98\arcsec$). However, for any physical $K_P-K$ (i.e., a star that is sufficiently blue and faint would land outside the galaxy) the fainter star is too faint to account for, or significantly dilute the transit depth of K2-100b, and was therefore ignored.

\section{Stellar Parameters}\label{sec:params} 

{\it Common parameters for all targets:} for each of the seven planet hosts, we assume a common [Fe/H], reddening, distance, and age. We adopt [Fe/H]=0.14$\pm$0.04, which encompasses measurements from the literature derived from high-resolution spectra of FGK stars \citep{Boesgaard2013, Yang2015, Netopil2016} and any chemical inhomogeneities as observed in similar clusters \citep[$\simeq0.02$~dex in Hyades,][]{Liu2016}. We adopt a reddening of $E(B-V)=0.027\pm0.004$~mag measured by \citet{Taylor2006}. Because the cluster is relatively compact ($1\sigma \simeq$2\,pc), and cluster and circumstellar gas (and dust) have been dispersed, reddening variation between targets should be smaller than measurement uncertainties. \citet{van-Leeuwen2009} measure a {\it Hipparcos}-based distance of 181.5$\pm6.0$\,pc to the core of Praesepe, which we use for each of the individual stars. As with reddening, the distance error is large enough to account for the scatter in individual object distances due to the finite size of the cluster core. This distance is consistent with independent measurements for Praesepe \citep[e.g.,][]{2009ApJ...697.1578G}. Main-sequence turnoff and isochrone fitting suggest an age for Praesepe and Hyades of 600-700\,Myr \citep{1998A&A...331...81P,2004A&A...414..163S}, but accounting for the effects of rotation and revisions to the solar metallicity scale suggest an older age of $\simeq$800\,Myr \citep{Brandt2015}. For our analysis, we adopted the older age of 790$\pm$30\,Myr from \citet{Brandt2015}. Global parameters are summarized in Table~\ref{tab:global}.

\begin{deluxetable}{l l l l}
\tabletypesize{\scriptsize}
\tablecaption{Global Parameters \label{tab:global}}
\tablewidth{0pt}
\tablehead{
\colhead{Parameter}&\colhead{Value}&\colhead{Source}\\
}
\startdata
Age (Myr) & 790$\pm$30 Myr & \citet{Brandt2015} \\
\hline
& &  \citet{Boesgaard2013}, \\
\rm{[Fe/H]} & 0.14$\pm$0.04 & \citet{Yang2015},\\ 
& & \citet{Netopil2016} \\
\hline
E(B-V) (mag)& 0.027$\pm$0.004 & \citet{Taylor2006} \\
Distance (pc) & 181.5$\pm$6.0 & \citet{van-Leeuwen2009} \\
\enddata
\end{deluxetable}

{\it Radial Velocities:} RVs were determined from the IGRINS data as explained in \citet{Mann2016a} and G. N. Mace et al. (in preparation). To briefly summarize, we used telluric lines to improve the wavelength solution and provide a fixed zero-point across all observations. We then cross-correlated each IGRINS spectra against 150-230 spectra of RV standards with similar spectral types to the target. The final assigned RV and error is the robust mean and standard error of the cross-correlation across all templates. For targets with multiple measurements, we used the weighted mean of the measurements. For absolute RVs, errors are limited by the zero-point error of $153$\,\mps, which is due to limits on the RV precision of the templates.

{\it Membership in Praesepe:} all planet hosts are included in the Praesepe membership catalog of \citet{Kraus:2007yq} with membership probabilities of $\ge97\%$. However, \citet{Kraus:2007yq} calculations use only proper motions and photometry. RVs derived from our IGRINS spectra enable the calculation of more precise, three-dimensional probabilities. We first measured each target's photometric distance by comparing available optical and NIR photometry against the solar metallicity isochrones from \citet{Dotter2008}. We drew position and proper motion information from UCAC4 \citep{UCAC4} or SDSS \citep{Ahn:2012kx}, where available. We then computed Galactic $UVW$ kinematics by combining these data with the RVs. Membership probabilities were calculated following the Bayesian framework of \citet{2011MNRAS.416.3108R}. We drew Praesepe $UVW$ kinematics from \citet{van-Leeuwen2009} and values for field stars from \citet{Malo2013}. We selected a membership prior equal to the ratio of the number of stars in Praesepe to the number of field stars in the same region of the sky. To this end, we constructed a Praesepe CMD from APASS or SDSS $r$ $-$ 2MASS $K$ color of Praesepe members identified by \citet{Kraus:2007yq}. We considered all stars with $r<17$, within 8$^\circ$ of the Praesepe core, and 5$\sigma$ of the Praesepe color-magnitude diagram (CMD); those not in the \citet{Kraus:2007yq} catalog, we assigned as field stars and those in \citet{Kraus:2007yq} we assigned as members, ignoring the individual membership probabilities for simplicity. The resulting Bayesian membership probabilities were $>99.9\%$ for all seven planet hosts. 

\begin{figure}
	\centering
	\includegraphics[width=0.43\textwidth]{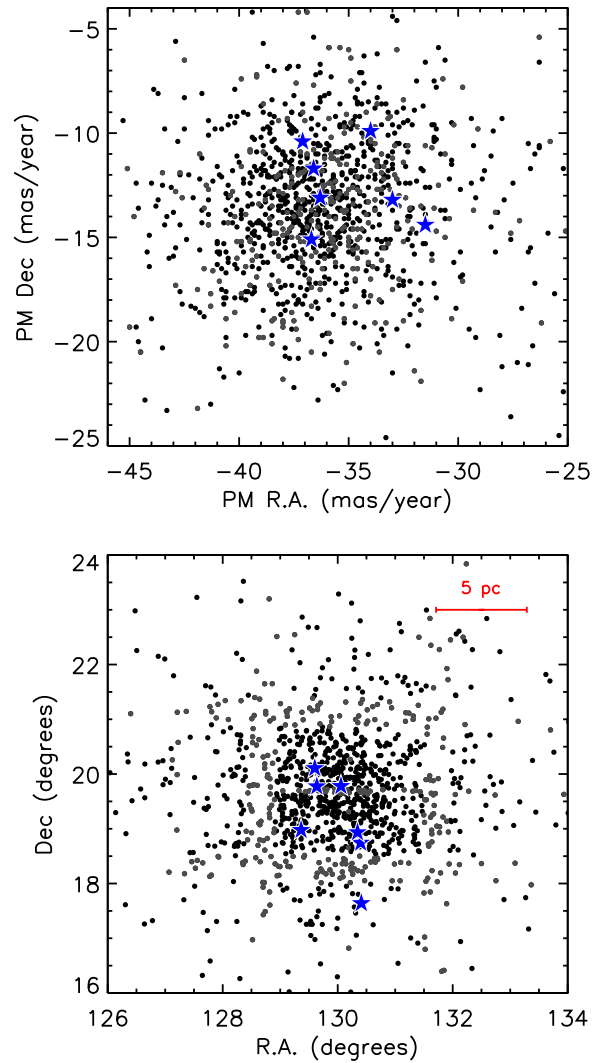} 
	\caption{Proper motions (top) and positions (bottom) of likely Praesepe members from \citet{Kraus:2007yq}. An estimate of the (projected) physical scale is shown in the center plot based on the distance to the cluster. Targets not observed by \ktwo\ are shown in gray. Planet hosts are shown as blue stars. Plot edges cut off some ($<5\%$) members to better show detail in the core. }
	\label{fig:praesepe}
\end{figure}

\begin{figure*}
	\centering
	\includegraphics[width=0.95\textwidth]{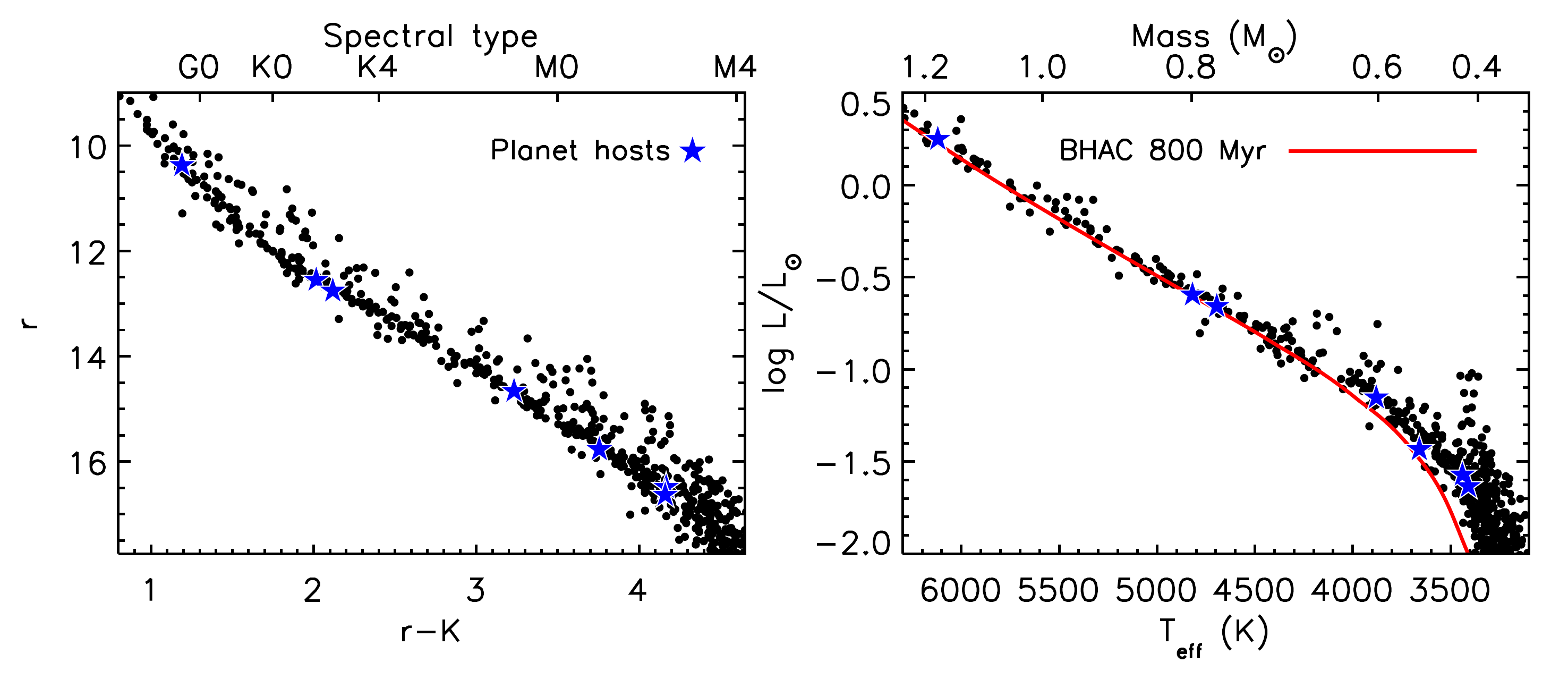} 
	\caption{Color--magnitude (left) and H--R diagram (right) for planet hosts (blue stars) and likely Praesepe members (black points) from \citet{Kraus:2007yq}, with an 800\,Myr isochrone from \citet{2015A&A...577A..42B} in red on the right panel. Approximate spectral types (left) and masses (right) are shown in the top axes. The bluest and reddest stars are cut from both plots to show more detail in the sequence near the planet hosts, and some stars are not shown in one or both plots due to a lack of sufficient reliable photometry.}
	\label{fig:praesepe_cmd}
\end{figure*}

RVs have not been measured for the majority of cluster members (especially low-mass members), so instead of $UVW$ we show positions and proper motions of planet hosts and cluster members in Figure~\ref{fig:praesepe}, and CMD positions in Figure~\ref{fig:praesepe_cmd}. All planet hosts are consistent with the kinematics and position of the cluster, and all planet hosts have CMD positions consistent with the single-star cluster sequence.

{\it Effective temperatures}: we compared our dereddened spectra to a grid of BT-SETTL CIFIST stellar atmosphere models\footnote{\href{https://phoenix.ens-lyon.fr/Grids/BT-Settl/CIFIST2011}{https://phoenix.ens-lyon.fr/Grids/BT-Settl/CIFIST2011}} \citep{2012RSPTA.370.2765A}. For the four M-dwarfs in the sample, we mask out regions where models poorly reproduce observed spectra as detailed in \citet{Mann2013c}, and for all stars we mask out regions of high telluric contamination. We included five nuisance parameters to deal with small errors in the wavelength and flux calibration of SNIFS and SpeX \citep[see ][ for more details]{Gaidos2014,Mann2015b}. We estimated errors due to uncertainties in the reddening by repeating the fitting process over the range of $E(B-V)$ values ($0.027\pm0.004$) and find the change in \teff\ is negligible compared to other errors. 

Our model-fitting method reproduces M-dwarf temperatures measured from long-baseline optical interferometry \citep{Boyajian2012,Mann2013c}. For the three warmer stars we tested our method using a similar sample FGK dwarfs with interferometric temperatures from \citep{Boyajian:2012fk}, which we also used to estimate the errors on our method and test for systematic offsets. Our final \teff\ values for the planet hosts are also consistent with those derived from color-\teff\ relations from  \citet{Mann2015b} for the M-dwarfs and from \citet{2005ApJ...626..465R} and \citet{Pinsonneault:2012lr} for FGK dwarfs.

We estimated \teff\ for other cluster members using available APASS, 2MASS, and Tycho-2 \citep{Hog2000} photometry of likely members and the color-\teff\ relations from \citet{Mann2015b} for the M-dwarfs and from \citet{2005ApJ...626..465R} and \citet{Pinsonneault:2012lr} for FGK dwarfs. We show the resulting \teff\ values with luminosities (see below) for planet hosts and members in Figure~\ref{fig:praesepe_cmd}. 

{{\it Bolometric fluxes}: we compiled well-calibrated photometry from the literature; $BVgri$ from the ninth data release of the AAVSO All-Sky Photometric Survey \citep[APASS,][]{Henden:2012fk}, $JHK_S$ from The Two Micron All Sky Survey \citep[2MASS,][]{Skrutskie2006}, $griz$ from the Sloan Digital Sky Survey \citep[SDSS,][]{Ahn:2012kx}, and $W1W2W3$ photometry from the Wide-field Infrared Survey Explorer \citep[WISE,][]{Wright:2010fk}. We then scaled the (still reddened) NIR and optical spectrum to the archival photometry following the procedure from \citet{Mann2015b}, including filling in regions outside our observed spectra (0.35--2.4\um) and areas of high telluric absorption with an atmospheric model. To calculate the bolometric flux (\fbol), we removed the effects of extinction/reddening from the combined and calibrated spectrum using the reddening law from \citet{1989ApJ...345..245C}, and then integrated the spectrum over all wavelengths. As with \teff, we repeated our routine using the range of possible $E(B-V)$ values, which had only a marginal effect on our overall \fbol\ errors. 

{\it Stellar radius, luminosity, and mass}: combining \fbol, \teff, and distance, we calculated stellar radii using the Stefan--Boltzman relation. We similarly computed luminosities from the distance and \fbol. We derived un-reddened, synthetic $K$ and $V$ magnitudes from our calibrated spectra using the filter profiles and zero-points from \citet{2003AJ....126.1090C} and \citet{Mann2015a}, which we converted to absolute magnitudes using the cluster distance. With these, we estimated host star masses using the the semi-empirical mass--$M_K$ relation from \citet{Mann2015b}, which reproduces the mass--radius relation from low-mass eclipsing binaries \citep{Feiden2012a}, and mass--luminosity relation from astrometric binaries \citep{Delfosse2000}. For the warmer stars, we use the empirical mass--$M_V$ relation from \citet{Henry:1993fk}. As a check, we also derive masses by interpolating $M_J$, $M_H$, and $M_K$ onto 800\,Myr stellar isochrones from \citet{2015A&A...577A..42B}. The model-based masses are all within 1$\sigma$ of those estimated above, but give errors that may be unrealistically small due to systematic errors in the underlying models, so we adopt the more empirical values.

Luminosities for the cluster population were estimated using the cluster distance and $r$-band bolometric corrections from \citet{Mann2015b} or $V$-band bolometric corrections from \citet{2003AJ....126..778V}. We excluded FGK stars lacking a $V$ magnitude and M-dwarfs lacking an $r$ magnitude. We use these luminosities and our \teff\ values (above) to create an H-R digram of the cluster, including the planet hosts, which we show in Figure~\ref{fig:praesepe_cmd} alongside the model H-R diagram from \citet{2015A&A...577A..42B}. 

{\it Rotation periods}: following the procedures described in \citet{Gaidos:2017aa}, we attempted to compute rotation periods for 908 Praesepe candidates observed by \ktwo. We downloaded K2SFF light curves, which we normalized and fit and subtracted off a second-order polynomial fit to each using robust methods. Rotation periods were calculated using both a Lomb--Scargle periodogram \citep{Scargle1981} and the autocorrelation function (ACF). The ACF value, if available, was preferred over the Lomb--Scargle value because the former is more robust to changes in light curve shape \citep{McQuillan:2013}.  Because the distributions of star spots often generate a second harmonic of the rotation period in the light curve, only the first and second peaks in the ACF were considered, and the higher of the two peaks was selected as the rotation period. To obtain a refined estimate of the period, a Gaussian function was fit to the ACF around the peak. Rotation periods $<35$d were successfully estimated for 738 stars, including all seven host stars; other stars exhibited no significant periodic variability or an ambiguous period. We exclude another 37 stars due to questionable $r$ or $K$ magnitudes. From the remaining 701 stars, 27 have independent rotation period measurements from \citet{2011ApJ...740..110A}, only one of which differs from our own measurements by more than expected errors. We show the distribution of rotation periods of the 701 stars with mass and color in Figure~\ref{fig:rot}.    

\begin{figure*}[htb!]
	\centering
	\includegraphics[width=0.9\textwidth]{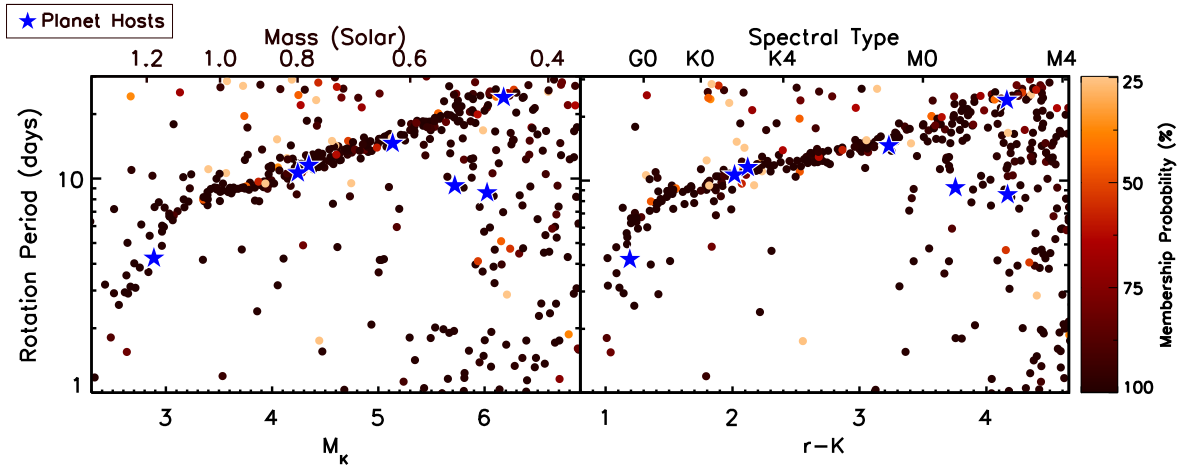} 
	\caption{Rotation periods of likely Praesepe members drawn from \citet{Kraus:2007yq} as a function of $M_K$ (left) and $r-K$ color (right). Approximate stellar masses (left) and spectral types (right) are shown on the top axes. Points are color-coded by their membership probabilities (most of which are $\simeq$100\%). The planet hosts are shown as blue stars, all of which have $\simeq$100\% membership probabilities. Rotation periods are determined from an autocorrelation function, or Lomb--Scargle periodogram where autocorrelation fails. The sequence is relatively tight for $M_K\lesssim5.5$, $r-K\lesssim3.5$; many of the outliers in this range are likely due to binarity, non-member interlopers, differential rotation, and/or poor detection or detection of the alias of the true rotation period \citep[e.g.,][]{2016ApJ...822...47D}.}
	\label{fig:rot}
\end{figure*}

{\it \vsini:} projected rotational velocities were measured using our high-resolution IGRINS data. We obtained a BT-SETTL model spectrum for each target consistent with the stellar parameters derived above. The model spectra were broadened by a Gaussian to match the instrument profile, which we measured from width of the telluric lines, as described in \citet{Mann2016a}. For each IGRINS order with S/N$>20$, we simultaneously fit for \vsini\ and five other nuisance parameters to handle flux calibration, wavelength calibration, and imperfectly corrected telluric lines. We assumed a microturbulent velocity of 1.5 \kms\ for all stars, and linear limb-darkening coefficients derived from PHOENIX models (see Section~\ref{sec:transitfit}) matching the parameters of each target. After fitting, we identify large $\gg5\sigma$ outliers in the residuals, which we mask out and repeat the process. Most of these outlier regions appear to be missing/erroneous lines in the model spectra or poorly corrected tellurics. 

For our final \vsini\ we adopted the robust weighted mean and standard error across all orders. Based on a comparison between our own measurements of \vsini\ and those in the literature of the same young stars \citep[e.g.,][]{Mermilliod2009,Dahm2012} we add an additional systematic error of 0.6 \kms. We attribute this extra error to poorly understood systematics, such as micro- or macroturbulence, improperly corrected instrumental broadening, and imperfect limb-darkening corrections. Because of the limiting resolution of the spectrograph, we consider measurements below 2.6\,\kms\ to be upper limits, which is the case for three targets. Based on their radii and rotation periods, all three have equatorial velocities below this limit, so only upper limits are expected.

{\it Sky projected stellar inclination}: $i_*$ can be compared to the planetary inclination ($i$) measured from the transit (usually $\simeq90^{\circ}$) as a probe of the planetary spin-orbit alignment. This in turn can be used to probe the formation or dynamical history of the planetary system. 
While this method is not as accurate as measurements from asteroseismology \citep[e.g.,][]{2013Sci...342..331H} or Rossiter-McLaughlin \citep[e.g.][]{2010PASJ...62L..61N}, it is still sufficient to identify highly misaligned systems and can provide meaningful constraints when applied to populations of planet hosts \citep[e.g.,][]{2013MNRAS.436.1883W}. Furthermore, \vsini\ and $P_{\rm{rot}}$, and therefore $i_*$ are generally more easily measured in rapidly rotating stars like those in this study. 

We calculate $i_*$ from \vsini, $P_{\rm{rot}}$, and $R_*$ using the formalism from \citet{Morton2014b}. For targets with only upper limits on \vsini\ we do not attempt to derive $i_*$. We ignored effects from differential rotation. To handle regions where \vsini$>V_{\rm{eq}}$ (which is unphysical), we converted \vsini\ and $V_{\rm{eq}}$ to a posterior in $cos(i_*)$. In all other cases, the resulting $i_*$ posteriors only provide lower limits on $i_*$ because they are all consistent with spin-orbit alignment. 

A summary of all derived stellar parameters and errors is given in Table~\ref{tab:stellar}.

\floattable
\begin{deluxetable*}{l | l l l l l l l l l l }
\tablecaption{Stellar Parameters}
\setlength{\tabcolsep}{3pt}
\tablewidth{0pt}
\tablehead{
\colhead{Parameter} & \colhead{K2-100} & \colhead{K2-101} & \colhead{K2-102} & \colhead{K2-103} & \colhead{K2-104} & \colhead{EPIC 211901114} & \colhead{K2-95} &
}
\startdata
\hline
EPIC \# & 211990866 & 211913977 & 211970147 & 211822797 & 211969807 & 211901114 & 211916756   \\
$\alpha$ R.A. (hh:mm:ss) & 08:38:24.302 & 08:41:22.581 & 08:40:13.451 & 08:41:38.485 & 08:38:32.821 & 08:41:35.695 & 08:37:27.058 \\
$\delta$ Dec. (dd:mm:ss) & +20:06:21.83 & +18:56:01.95 & +19:46:43.72 & +17:38:24.02 & +19:46:25.78 & +18:44:35.01 & +18:58:36.07 \\
$\mu_{\alpha}$ (mas~yr$^{-1}$) & $ -35.7\pm 0.6$ & $ -34.3\pm 1.8$ & $ -37.1\pm 3.0$ & $ -36.4\pm 2.5$ & $ -34.7\pm 3.9$ & $ -34.0\pm 3.0$ & $ -36.0\pm 3.0$ \\
$\mu{\delta}$ (mas~yr$^{-1}$) & $ -13.1\pm  0.6$ & $  -9.6\pm  2.1$ & $ -14.3\pm  2.0$ & $ -11.8\pm  2.7$ & $  -6.5\pm  4.0$ & $ -11.0\pm  3.0$ & $ -13.0\pm  3.0$ \\
$\mu$ source & UCAC4 & UCAC4 & UCAC4 & UCAC4 & UCAC4 & SDSS & SDSS \\
$r$ (mag) & $10.373\pm0.048$ & $12.552\pm0.036$ & $12.758\pm0.020$ & $14.661\pm0.004$ & $15.770\pm0.004$ & $16.485\pm0.005$ & $16.635\pm0.006$ \\
$r$ Source & APASS & APASS & APASS & SDSS & SDSS & SDSS & SDSS \\
$J$ (mag) & $  9.46\pm  0.03$ & $ 11.16\pm  0.02$ & $ 11.28\pm  0.02$ & $ 12.28\pm  0.03$ & $ 12.88\pm  0.03$ & $ 13.15\pm  0.02$ & $ 13.31\pm  0.02$ \\
$H$ (mag) & $  9.24\pm  0.03$ & $ 10.68\pm  0.02$ & $ 10.74\pm  0.02$ & $ 11.61\pm  0.03$ & $ 12.25\pm  0.02$ & $ 12.54\pm  0.02$ & $ 12.74\pm  0.02$ \\
$K_S$ (mag) & $  9.18\pm  0.02$ & $ 10.54\pm  0.02$ & $ 10.64\pm  0.02$ & $ 11.43\pm  0.02$ & $ 12.01\pm  0.02$ & $ 12.32\pm  0.02$ & $ 12.47\pm  0.02$ \\
Rotation Period (days) & $  4.3\pm  0.1$ & $ 10.6\pm  0.6$ & $ 11.5\pm  0.7$ & $ 14.6\pm  1.1$ & $  9.3\pm  0.4$ & $  8.6\pm  0.4$ & $ 23.9\pm  2.4$ \\
Barycentric RV (\kms) & $33.60\pm0.30$ & $34.31\pm0.17$ & $34.85\pm0.17$ & $34.85\pm0.17$ & $34.81\pm0.17$ & $34.06\pm0.17$ & $35.85\pm0.17$ \\
\vsini\ (km~s$^{-1}$) & $14.8^{+0.8}_{-0.8}$ & $3.9^{+0.9}_{-0.7}$ & $3.0^{+1.0}_{-0.7}$ & $<2.6$ & $<2.6$ & $3.2^{+1.0}_{-0.7}$ & $<2.6$ \\
$i_*$ (degrees) & $>77$ & $>66$ & $>61$ & \nodata & \nodata & $>64$ & \nodata \\
\teff\ (K) & $6120\pm 90$ & $4819\pm 45$ & $4695\pm 50$ & $3880\pm 67$ & $3660\pm 67$ & $3440\pm 65$ & $3410\pm 65$ \\
$M_*$ ($M_\odot$) & $ 1.18\pm 0.09$ & $ 0.80\pm 0.06$ & $ 0.77\pm 0.06$ & $ 0.61\pm 0.02$ & $ 0.51\pm 0.02$ & $ 0.46\pm 0.02$ & $ 0.43\pm 0.02$ \\
$R_*$ ($R_\odot$) & $  1.19\pm 0.05$ & $  0.73\pm 0.03$ & $  0.71\pm 0.03$ & $  0.59\pm 0.03$ & $  0.48\pm 0.02$ & $  0.46\pm 0.02$ & $  0.44\pm 0.02$ \\
$L_*$ ($L_\odot$) & $ 1.777\pm0.062$ & $0.2542\pm0.0093$ & $0.2201\pm0.0082$ & $0.0703\pm0.0021$ & $0.0368\pm0.0012$ & $0.0268\pm0.0010$ & $0.0232\pm0.0009$ \\
$\rho_*$ ($\rho_\odot$) & $ 0.70^{+0.11}_{-0.09}$ & $ 2.07^{+0.29}_{-0.25}$ & $ 2.14^{+0.31}_{-0.27}$ & $ 2.98^{+0.43}_{-0.38}$ & $ 4.64^{+0.68}_{-0.60}$ & $ 4.62^{+0.69}_{-0.60}$ & $ 5.16^{+0.77}_{-0.67}$ \\
\hline
\enddata
\label{tab:stellar}
\tablecomments{All $JHK_S$ magnitudes are from 2MASS.}
\end{deluxetable*}

\section{Transit Fitting}\label{sec:transitfit}

We fit all \ktwo\ light curves with a Monte Carlo Markov Chain (MCMC) as described in \citet{Mann2016a}, which we briefly summarize here. We used the {\it emcee} Python module \citep{Foreman-Mackey2013} to fit the model light curves produced by the \textit{batman} package \citep{Kreidberg2015} using the \citet{MandelAgol2002} algorithm. Following \citet{Kipping:2010lr} we over-sampled and binned the model to match the 30\,minute \ktwo\ cadence. We sampled the planet-to-star radius ratio ($R_P/R_*$), impact parameter ($b$), orbital period ($P$), epoch of the first transit mid-point ($T_0$), bulk stellar density ($\rho_*$), two parameters that describe the eccentricity and argument of periastron ($\sqrt{e}\sin\omega$ and $\sqrt{e}\cos\omega$), and two (quadratic) limb-darkening parameters ($q1$ and $q2$).

We assumed a quadratic limb-darkening law and use the triangular sampling method of \citet{Kipping2013} in order to uniformly sample the physically allowed region of parameter space. We applied a prior on limb-darkening derived from the \citet{2013A&A...553A...6H} atmospheric models, calculated using the LDTK toolkit \citep{2015MNRAS.453.3821P}, which enabled us to account for errors in stellar parameters. For this, we used the filter and CCD transmission function for \kepler\ from the \kepler\ science center\footnote{\href{http://keplergo.arc.nasa.gov/CalibrationResponse.shtml}{http://keplergo.arc.nasa.gov/CalibrationResponse.shtml}} and stellar parameters and errors derived in Section~\ref{sec:params}. Errors on the limb-darkening coefficients were broadened to account for model uncertainties (estimated by comparing limb-darkening parameters from different model grids). Typical resulting errors on $u1$ and $u2$ were 0.08 and 0.04, respectively. 

For each system (excluding EPIC 211901114b, see below), we ran two MCMC fits. For the first, we fixed $e$ and $\omega$ to zero, and used a uniform prior on $\rho$, and for the second fit we allowed $\sqrt{e}\sin\omega$ and $\sqrt{e}\cos w$ to float from 0 to 1 under uniform priors and put a Gaussian prior on $\rho$ using our stellar parameters from Section~\ref{sec:params}. In both cases $\rho$ is forced to be $>0$, but has no upper bound. 

For EPIC 211901114b the transit duration is comparable to or less than the \kepler\ long-cadence integration time (30 min, see Figure~\ref{fig:211901114}). The light curve can therefore provide only an upper limit on transit duration (or a lower limit on stellar density for $e=0$) without additional constraints. To mitigate this, we fit the transit with $e$ and $\omega$ fixed at zero and simultaneously apply a Gaussian prior on $\rho$. Since the planet could be truly eccentric, assuming $e=0$ may bias the resulting fit parameters. However, even with the $e=0$ constraint, the final transit-fit parameters are still highly uncertain, so it is unavoidable if we want to make any inferences about the planet. We urge caution when interpreting the transit fit for this system. 

\begin{figure}
	\centering
	\includegraphics[width=0.95\columnwidth]{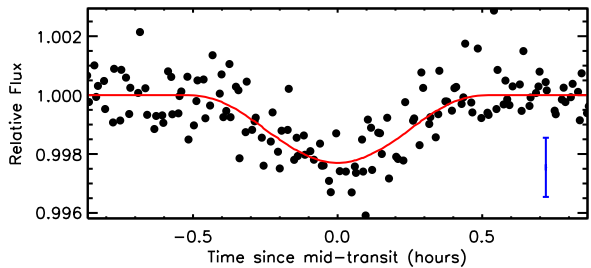} 
	\caption{Phase-folded light curve of EPIC~211901114b from \ktwo\ (black points). The best-fit (highest likelihood) transit model is shown as a red line and an estimate of the photometric errors on each point is shown in the bottom right in blue.}
	\label{fig:211901114}
\end{figure}

All MCMC fits were allowed to explore $|b|<1+R_P/R_*$, $P$ from 0 to 35 days, $R_P/R_*$ from 0 to 0.5, and $T_0$ within $P$/2 of the initial value, under uniform priors. For all targets, $R_P/R_*>0.5$ is conservatively ruled out by the lack of a second set of lines in our IGRINS spectrum and their locations on a color-magnitude diagram (Figure~\ref{fig:praesepe}). All parameters were initialized to the values from our BLS search (Section~\ref{sec:k2}), which are based on a Levenberg-Marquardt fit to the light curve \citep{Markwart2009}. MCMC chains were run using 150 walkers, each with 150,000 steps including a burn-in phase of 15,000 steps that was stripped from the final posteriors. Examination of the final posterior distribution suggests our selected numbers of steps, walkers, and burn-in length are more than sufficient for convergence.

\begin{figure*}
	\centering
	\includegraphics[width=0.95\textwidth]{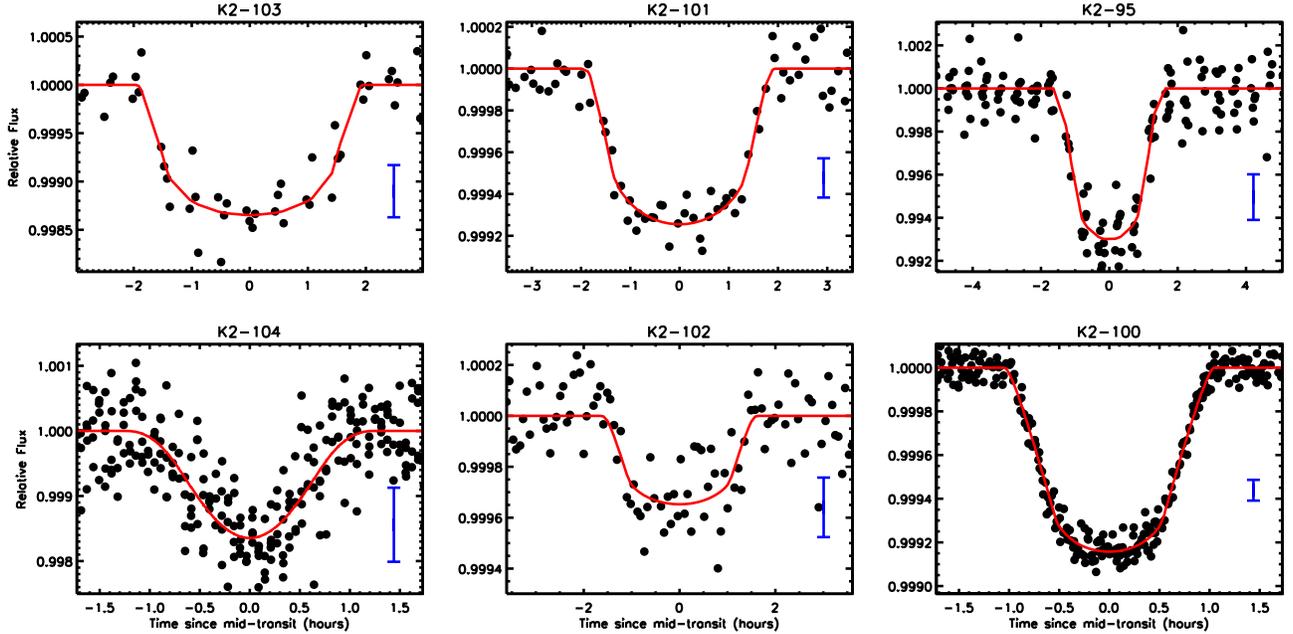} 
	\caption{Phase-folded light curve of six of the transiting planets from \ktwo\ (black points). The best-fit (highest likelihood) transit models are shown as red solid lines. Estimates of the photometric errors for each star are shown as blue error bars in the bottom right corner of each panel.}
	\label{fig:transitfit}
\end{figure*}

We report the transit-fit parameters in Table~\ref{tab:planet}. For each parameter, we report the median value with the errors as the 84.1 and 15.9 percentile values (corresponding to 1$\sigma$ for Gaussian distributions). The model light curves with the best-fit models for six of the seven systems shown in Figure~\ref{fig:transitfit}, with EPIC~211901114 shown in Figure~\ref{fig:211901114}. We also show the distributions and correlations for a subset of parameters ($\rho$, $e$, $b$, and $R_P/R_*$) in Figure~\ref{fig:correlations} (excluding EPIC 211901114b) with the median and statistical mode for each parameter marked.

\begin{figure*}
	\centering
	\includegraphics[width=0.43\textwidth]{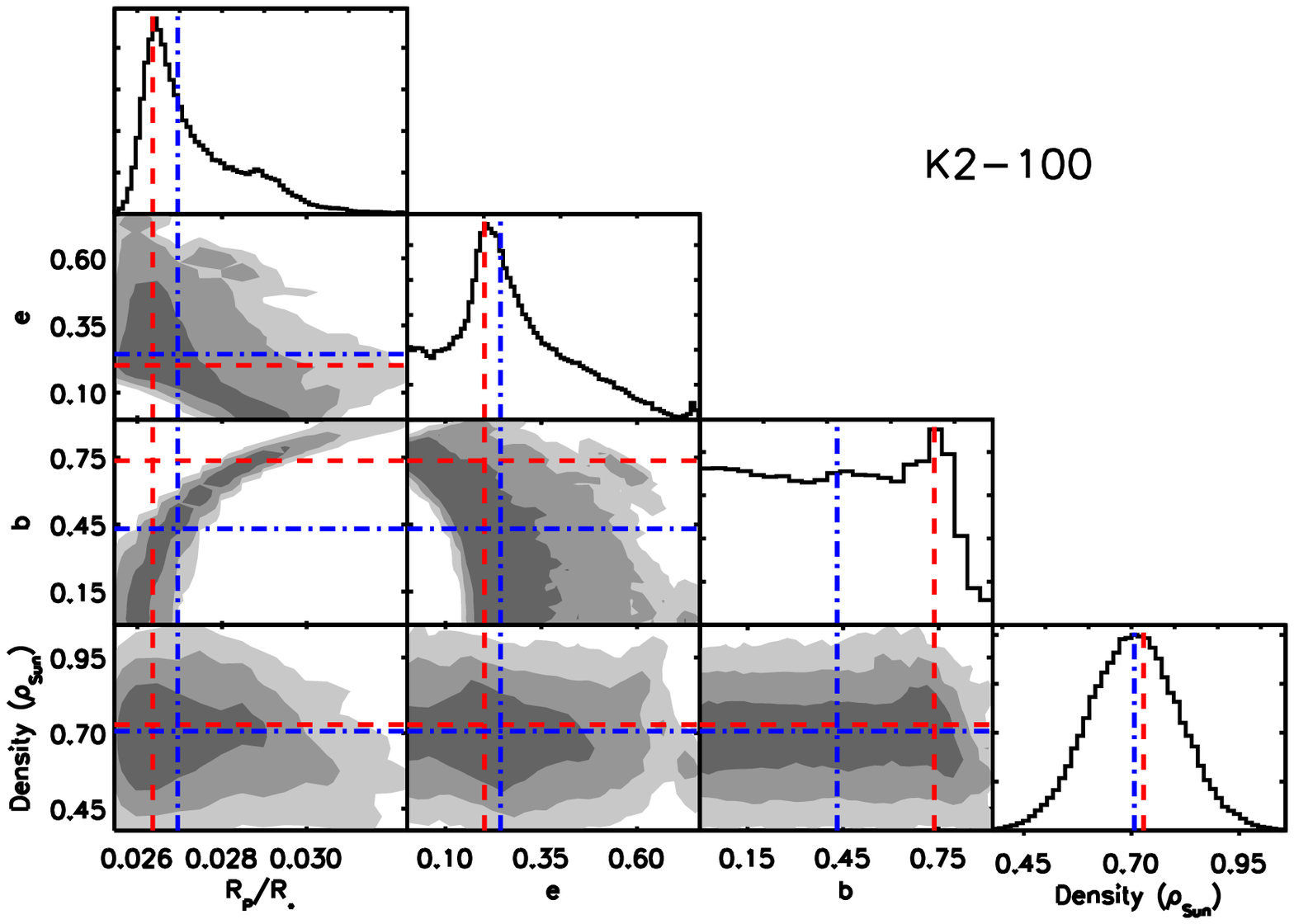} 
	\includegraphics[width=0.43\textwidth]{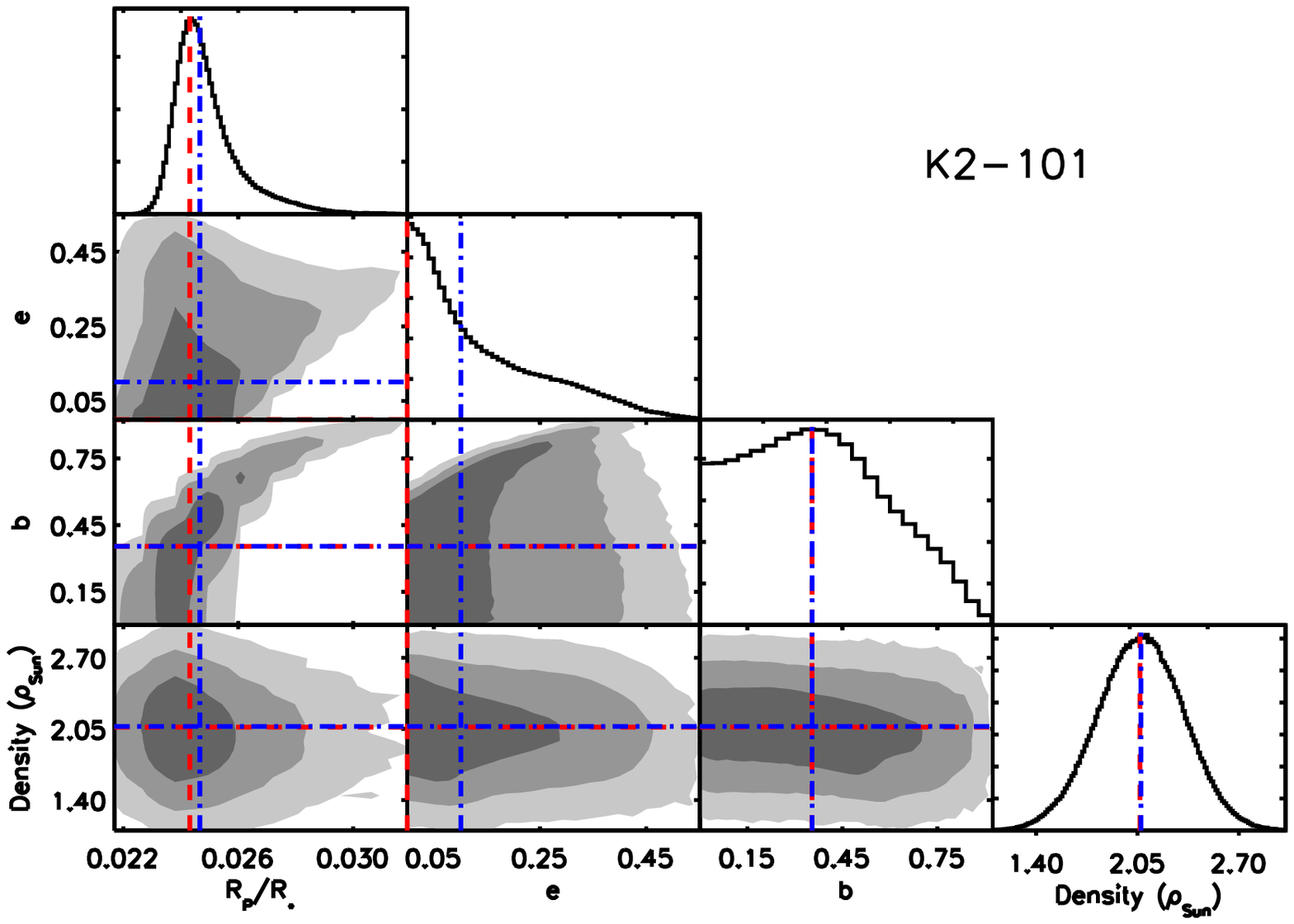} 
	\includegraphics[width=0.43\textwidth]{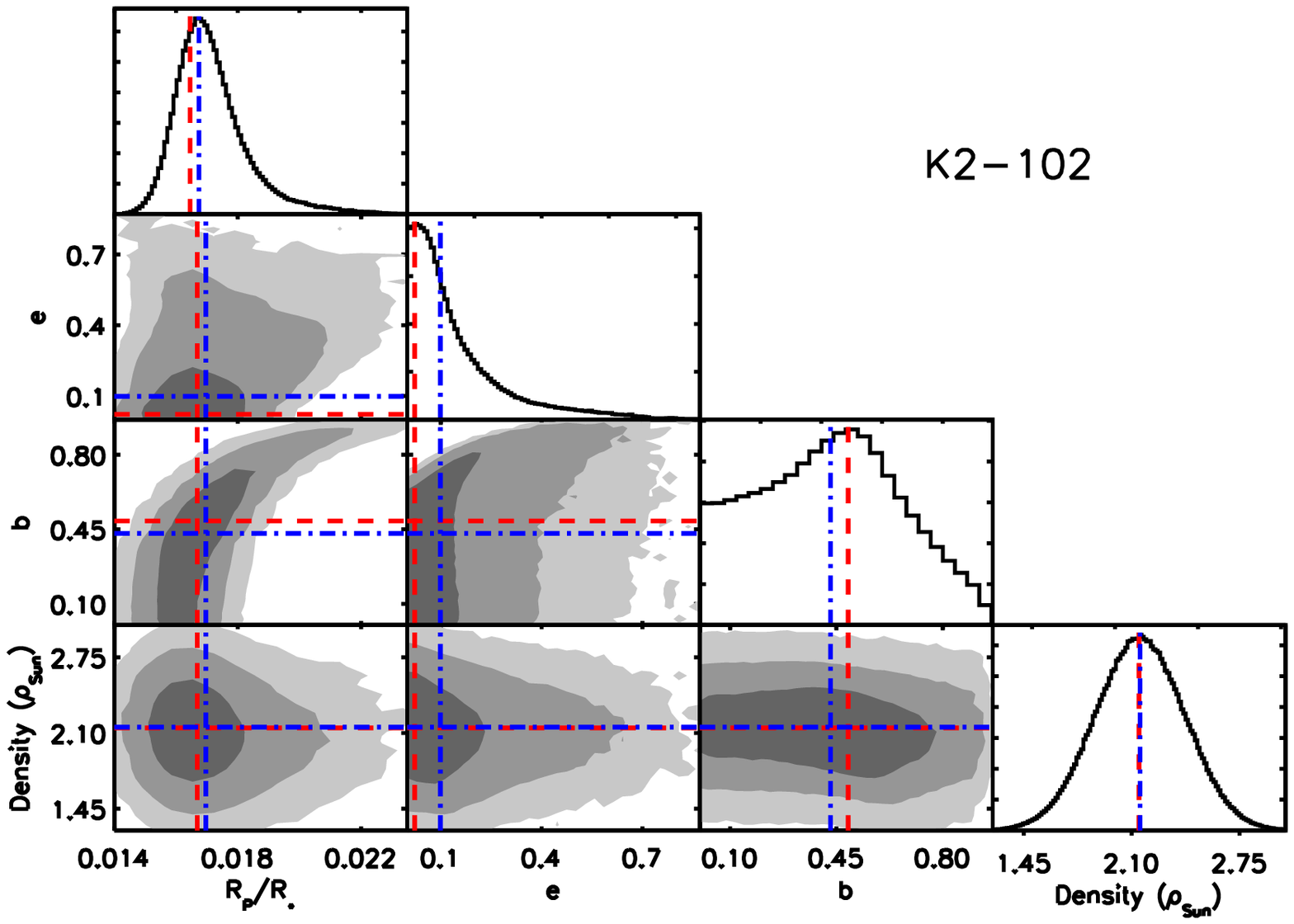} 
	\includegraphics[width=0.43\textwidth]{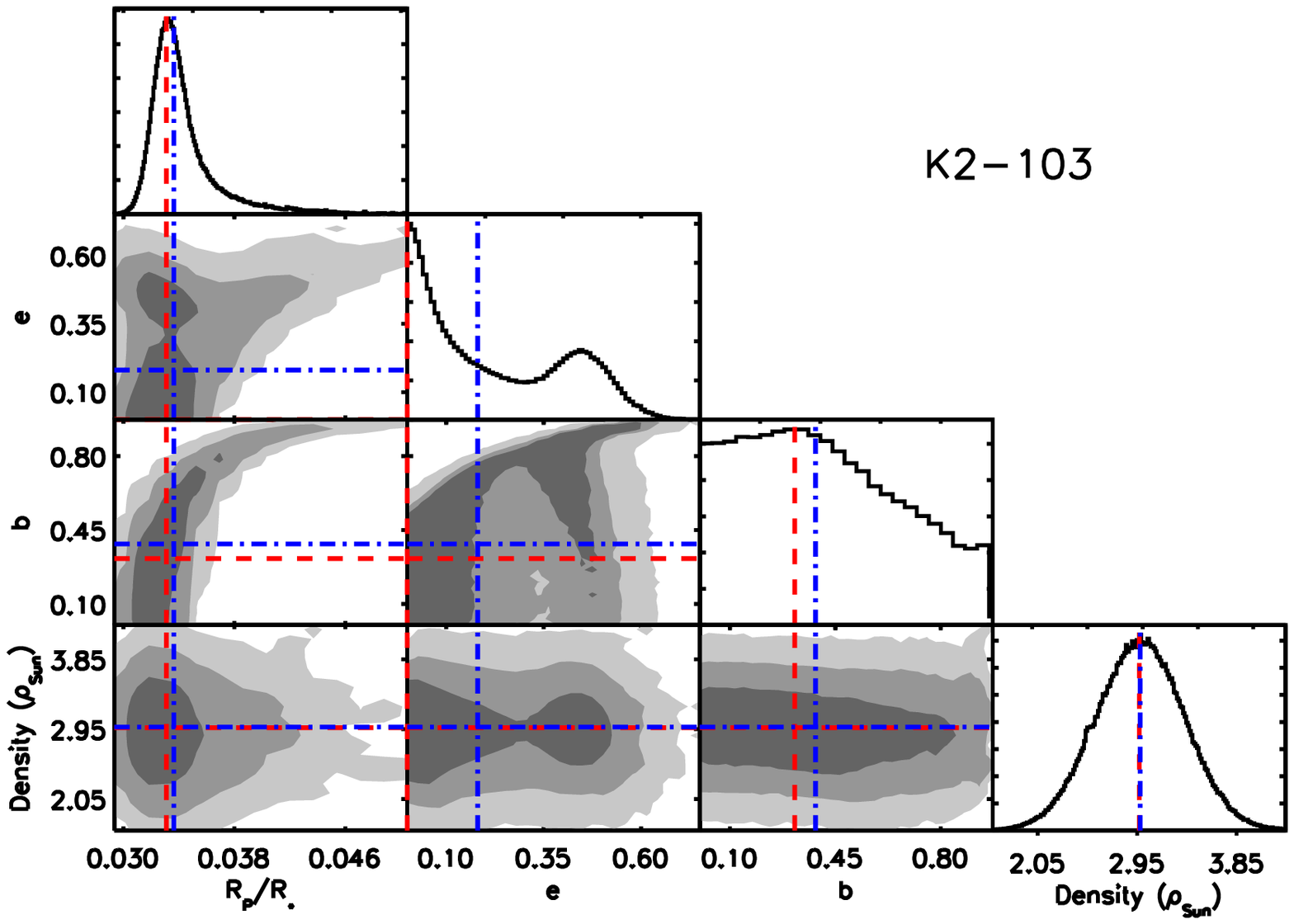} 
	\includegraphics[width=0.43\textwidth]{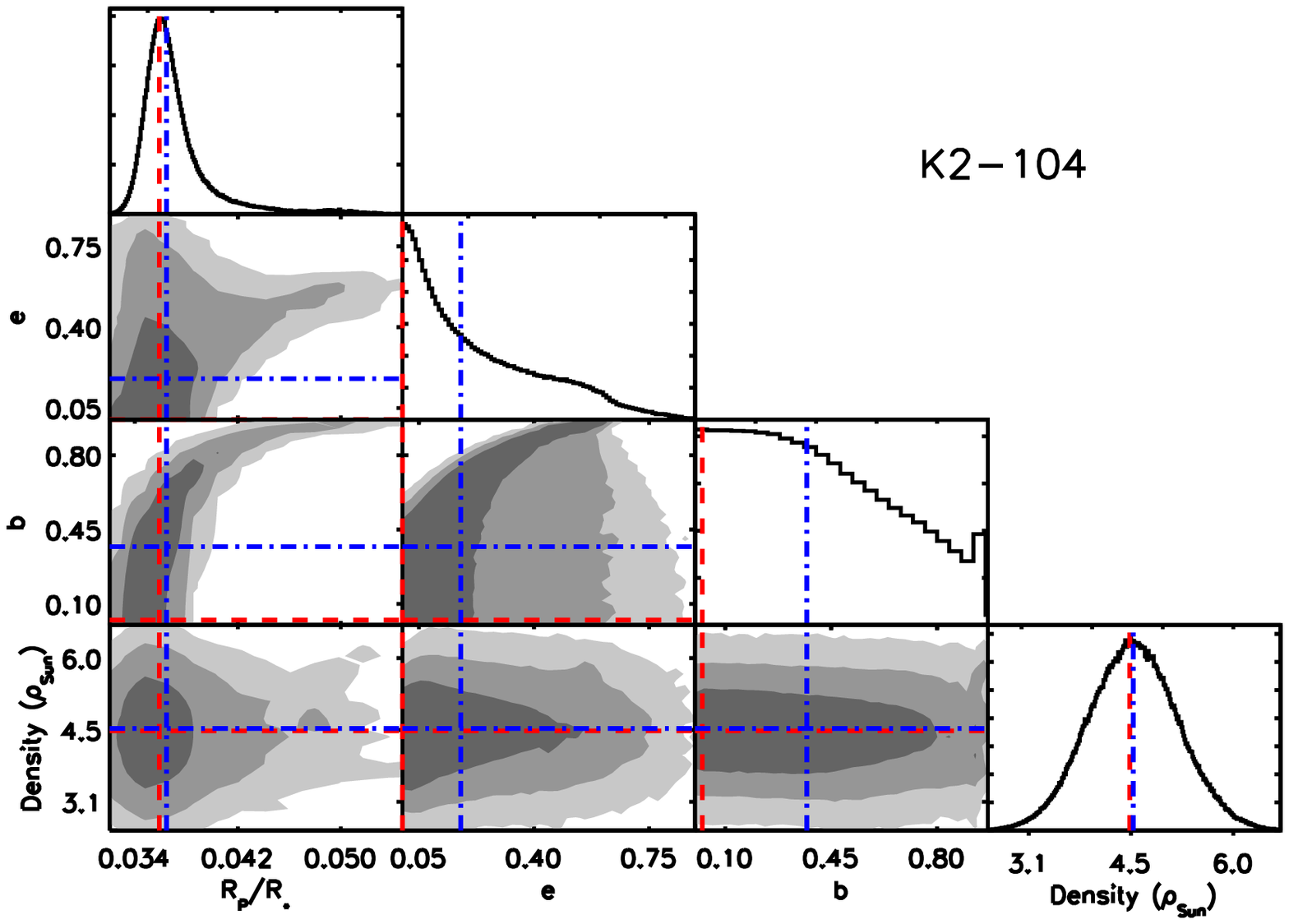} 
	\includegraphics[width=0.43\textwidth]{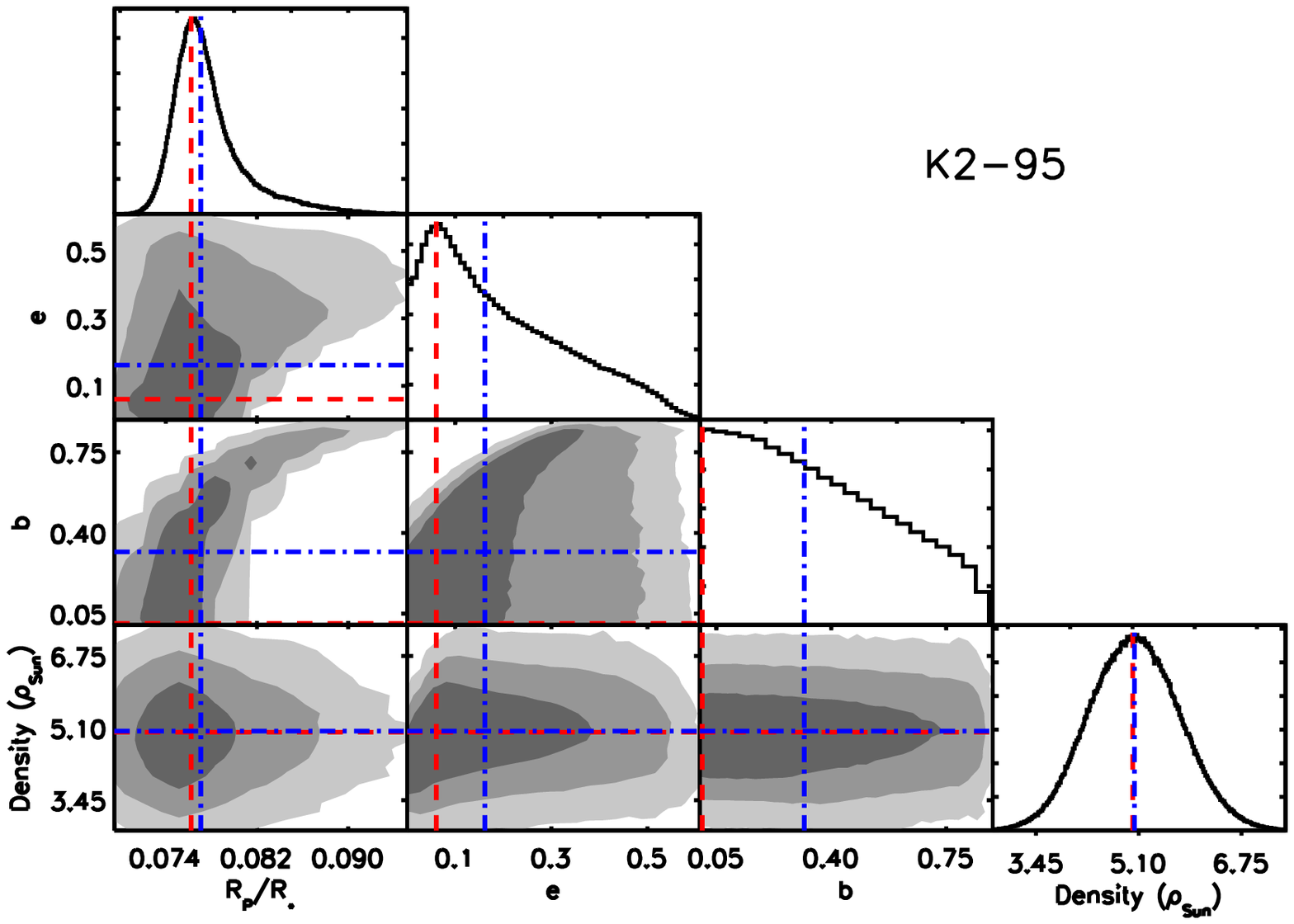} 
	\caption{Distributions and correlations between $\rho$, $e$, $b$, and $R_P/R_*$ for the six systems where $e$ was allowed to float (so EPIC 211901114 is excluded). The gray regions contain 68\%, 95\%, and 99.7\% of the points (from darkest to lightest). MCMC steps with $b<0$ are allowed in our MCMC, but the posteriors are symmetric about $b=0$ so we instead show $|b|$. The red dashed lines mark the statistical mode and the blue dotted-dashed blue lines correspond to the median of each distribution. Plot ranges exclude a small fraction of the points ($<1\%$), so more clarity can be seen in the main distribution. }
	\label{fig:correlations}
\end{figure*}

\section{False-positive Analysis}\label{sec:fpp} 
We estimated the likelihood that a given candidate is a true planet using the \textit{vespa} software \citep{2012ApJ...761....6M, 2015ascl.soft03011M}. \textit{vespa} considers three astrophysical false-positive scenarios; background eclipsing binaries, bound eclipsing binaries, hierarchical eclipsing systems, and each of the three but at double the reported period. \textit{vespa} then compares the likelihood of each false-positive scenario to that of a planet accounting for the shape and depth of the transit, the properties of the star, and external constraints from our AO imaging (where available). All FPP values are listed in Table~\ref{tab:planet}. 

With the exception of EPIC~211901114b, all candidates are assigned false-positive probabilities (FPP) of $<1\%$, effectively confirming their planetary nature. EPIC~211901114b was initially assigned an ambiguous FPP of 36\%, owing primarily to a high probability of being an eclipsing binary (34\%). This is consistent with own transit-fit posterior, which does not rule out a stellar or brown dwarf radius ($>11R_\earth$). Our IGRINS-derived radial velocity measurements are consistent with no variation, ruling out any companion $>5$Jupiter masses at 5-$sigma$ at the planet candidate's orbital period and assuming a circular orbit (Figure~\ref{fig:rv}). This reduces the FPP to (2\%); the remaining FPP is due to the possibility that the system is a hierarchical eclipsing binary. The IGRINS spectra show only one set of lines and there is no evidence of an unresolved binary in the CMD position of EPIC~211901114, but these cannot rule out a companion eclipsing binary significantly fainter than the primary ($\Delta K>3$\,mags). We conservatively consider EPIC~211901114b unconfirmed, pending additional AO observations or higher cadence transit observations. 

\begin{figure}
	\centering
	\includegraphics[width=0.45\textwidth]{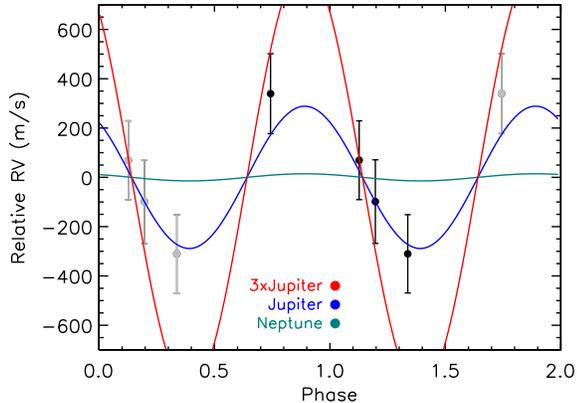} 
	\caption{Radial velocities for EPIC~211901114 phased to the candidate planet's orbital period. Gray points are repeated points. The expected variation due to Neptune, Jupiter, and 3xJupiter mass planets are shown as teal, blue, and red lines, all assuming circular orbits. }
	\label{fig:rv}
\end{figure}

\textit{vespa} does not consider false-positives due to instrumental correlated (`red') noise, which may be significant for \ktwo, nor does it consider stellar variability (spots), which is non-negligible for such young stars. We reject the first scenario for a number of reasons. First, with the exception of EPIC 211901114, all light curves show a transit-like shape, with a visible limb-darkened ingress, egress, and flattening bottom. The bottom of K2-104's light curve appears more V-shaped due to a short duration and significant limb-darkening, but this is consistent with expectations given the period and stellar parameters. Furthermore, none of the planets have orbital periods consistent with an alias of the \ktwo\ drift or thruster fire timescale. Lastly, all planets are detected in at least two of the publicly available reduced \ktwo\ light curves: K2SC \citep{2016MNRAS.459.2408A}, EVEREST \citep{Luger:2016aa}, and/or K2SFF \citep{Vanderburg2014}, and all are detected in our own extracted curves. This consistency suggests that no signal is an artifact of the data reduction process. 
 
We similarly reject the possibility that these signals are due to stellar variability. Flaring can be seen on some host light curves, including near transit, but all transits are detected even when data points near stellar flares are removed. The transit durations and shapes are inconsistent with any reasonable spot pattern. Most importantly, no planet has a period consistent with an alias of the measured rotation period, indicating that the two signals are independent of each other. 

Another scenario not directly considered by our {\it vespa} analysis is a blended planet, i.e., a bound or background star with a transiting planet creating the signal. In the case of a bound companion, or background star that is a member of the cluster, the planet would still orbit a member of Praesepe, so we do not consider these false-positives. However, our derived planet parameters would be incorrect due to incorrect stellar parameters and significant uncorrected dilution from the primary star. For a non-cluster member as the transit source, analysis of {\it Kepler} planet candidates suggests cases of background transiting planets are intrinsically rare \citep[1-4\%][]{Fressin:2013qy}. In either case, the star with the planet (bound or background) would need to be similar in brightness to the target to reproduce the transit depth, and would likely be seen as a higher CMD position, a second set of lines in the IGRINS spectra, or a companion in the AO/NRM data if available. 

\section{Discussion}\label{sec:discussion} 

\subsection{The Dynamical State of Close-in Planets at $\simeq$800\,Myr}

All systems are consistent with zero or low eccentricities ($\simeq0.2$) and alignment with their host star's rotation. This matches findings for the Hyades planet \citep[K2-25b, ][]{Mann2016a, David2016}. Of the seven systems (including K2-25b, but excluding EPIC 211901114b) K2-100b has the transit-fit posterior most consistent with a non-zero $e$, but can be reconciled with a larger impact parameter (see Figure~\ref{fig:correlations}). Similarly, all \vsini\ measurements are within expectations (given uncertainties) for spin-orbit aligned systems. Even those with only upper limits on \vsini\ are expected to have equatorial velocities well below IGRINS resolution. 

Unfortunately, our eccentricity measurements are all quite coarse (typical errors of 0.1-0.2), making it difficult to rule out small but non-zero eccentricities. However, the distribution of $e$ with $\omega$ in the transit-fit posteriors suggests the underlying eccentricity distribution of all systems is smaller than when considering each system individually. The value of $\rho_*$ derived from a transit light curve assuming zero eccentricity versus letting $e$ float will differ by a factor that depends on $\omega$ and $e$ \citep{Seager:2003lr,2012MNRAS.421.1166K}:
\begin{equation}
\rho_*^{\rm{circ}} \simeq \left(\frac{1+e\sin\omega }{(1-e^2)^{1/2}}\right)^3 \rho_* .
\end{equation}
The factor in front of $\rho_*$ can be $\simeq1$ for $e\simeq0$ or when $\sin\omega  \simeq \frac{(1-e^2)^{1/2}-1}{e}$. Thus fits to the light curve of a transiting planet with $e_{\rm{true}}=0$ can often yield answers with $e_{\rm{measured}}\gg0$ at specific values of $\omega$, especially in the presence of red and white noise \citep{2006MNRAS.373..231P, Gazak:2012vn}. This effect can be seen in our own fit posteriors, of which we show an example in Figure~\ref{fig:e_omega}. A similar pattern is seen in all fits where $e$ is not fixed to zero: values of $e\gg0$ tend to be clustered around specific values of $\omega$ that keep the transit observables unchanged. For near-circular orbits $\omega$ should be $\simeq$random, so the true $e$ distribution is likely smaller than the combined posteriors imply. 

\begin{figure}[htbp]
\begin{center}
	\includegraphics[width=0.43\textwidth]{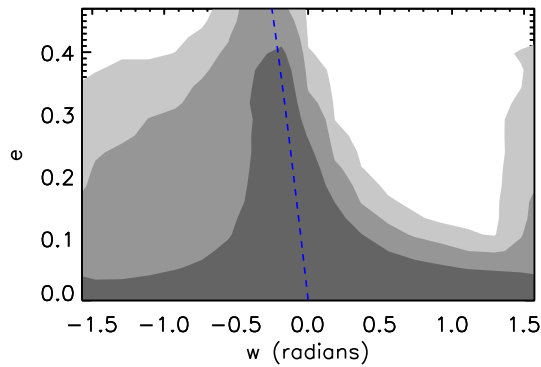} 
\caption{Correlation between $\omega$ and $e$ for K2-101b following the shading scheme of Figure~\ref{fig:correlations}. The blue dashed line marks where $\rho_*^{\rm{circ}}\simeq\rho_*$ and $e\ne0$, a region where large values of $e$ are allowed with minimal impact on the transit shape. }
\label{fig:e_omega}
\end{center}
\end{figure}

A solution to this is to generate simulated $e$ and $\omega$ distributions and compare them to the distribution of $\rho_{\rm{transit}}/\rho_*$ derived from our fits with $e=0$ and our stellar parameters from Section~\ref{sec:params}, similar to the procedure in \citet{Van-Eylen2015}. Assuming a Rayleigh distribution for $e$ and a uniform distribution in $\omega$, we can set an upper limit on the Rayleigh scale parameter ($\sigma_e$) of 0.26 at 95\%. A more sophisticated treatment would also account for the selection bias in favor of higher eccentricity systems \citep{2014MNRAS.444.2263K}, but this effect is smaller than current errors. Additional systems and higher cadence ground-based transit photometry would provide significantly improved constraints.

The spin-orbit alignment measurements, similar to measurements of $e$, provide only rough constraints. For the four systems where we detect \vsini\ broadening the measurements are consistent with alignment. Although \vsini\ is too small to detect in the other three systems, non-detection is expected based on their rotation periods and radii. Higher resolution observations would help with the missing systems, and measurements of the Rossiter-McLaughlin effect could provide significantly stronger constraints on spin-orbit alignment \citep[e.g.][]{2010PASJ...62L..61N}. In cases where the Rossiter-McLaughlin effect is too small to detect (due to small \vsini\ and transit depth) it may be possible to measure spin-orbit alignment by observing spot-crossing events \citep[e.g.,][]{2011ApJ...740L..10N}.

Studies of transiting planets around older ($\gtrsim 1$\,Gyr) stars from \kepler\ suggests Earth- to Neptune-size planets have small ($\lesssim0.1$) eccentricities \citep{2014ApJ...787...80H,Van-Eylen2015}, and their orbits are generally aligned with their host star \citep[e.g.,][]{2013MNRAS.436.1883W}. However, after dissipation of the protoplanetary gas disk, planet-planet scattering can drive super-Earth to Neptune size planets to large eccentricities ($\gg0.4$) and spin-orbit misalignment on $\sim$100\,Myr timescales \citep[e.g.,][]{2008ApJ...686..580C,2008ApJ...682.1264K}. While coarse, our findings suggest that such scattering or other highly disruptive events in young planetary systems are not the norm, and at 800\,Myr systems may be as dynamically settled as their old counterparts. 

Another tool used to probe the dynamical state of planetary systems is the number of multiplanet systems, which provides constraints on the level of mutual inclination of planets \citep[e.g.,][]{Ballard2016}. We detect no multiplanet system, yet approximately 22\% of \kepler\ systems are known to harbor multiple planets \citep{2014PNAS..11112647B}. This suggests that, of the 8 systems in Hyades and Praesepe we should have found 1.8 multiplanet systems, with a Poisson probability of 17\% of detecting none. If we cut the \kepler\ sample on S/N, period, and host star properties to simulate a \ktwo\ like survey of Praesepe ($P<35$\,days, fewer detected transits, only dwarf stars, etc.) then the fraction of multiples drops to 16\%, and the Poisson probability increases to 28\% of finding no multiplanet systems. So there is not a statistically significant deficit of multiples and we can draw no useful conclusions about the mutual inclination distribution of young planets at this time. 

\subsection{Why Are There So Few Planets in the Pleiades?}
There is a notable difference in the number of planets detected in the $\simeq$ 800 Myr old Praesepe and Hyades clusters (8 planets) versus in the $\simeq$ 125\, Myr old Pleiades cluster \citep[0 planets,][]{Gaidos:2017aa}. This is despite the fact that the target samples are similar in number ($\simeq$1000 targets in Praesepe + Hyades and $\simeq$1000 in Pleiades), spectral type distribution, and metallicity. Furthermore, Pleiades is closer than Praesepe (136\,pc versus 181\,pc), and hence similar mass stars are statistically brighter. It is possible that we are seeing signatures of planetary migration on 0.1-1\,Gyr timescales, but the differences are not yet statistically significant.

The difference in number of planets detected between the clusters may instead be due to faster and higher amplitude rotation for younger Pleiades stars, which can be difficult to remove and complicate the detection of short-period planets. A planet injection test assuming a \kepler-like population done by \citet{Gaidos:2017aa} indicates this is at least partially to blame for the lack of detections in Pleiades. Typical rotation periods in Pleiades are $1$\,d$<P_{\rm{rot}}<10$\,d for the range of spectral types probed by \ktwo\ \citep{2016ApJ...822...81C}. We detect three planets around stars with $P_{\rm{rot}}<10$\,days in Praesepe, but none around stars with $P_{\rm{rot}}<4$\,days (Figure~\ref{fig:rot}). However, many fast rotating stars ($P_{\rm{rot}}<4$\,days) in Praesepe are likely to be tidally locked binaries \citep{2016ApJ...822...47D}, around which close-in planets are significantly less common \citep{2014ApJ...791..111W,Kraus2016a}. So the lack of detections around the fastest rotators is not conclusive. 

A more detailed injection/recovery test across all nearby young clusters/star-forming regions (Upper Scorpius, Hyades, Pleiades, and Praesepe) to better constrain our completeness as a function of rotation period (and other stellar properties) would be useful, particularly when combined with the additional \ktwo\ observations of Hyades planned for Campaign 13. {\it TESS} is also expected to survey nearby young stars and clusters \citep{2014arXiv1410.6379S} and could shed significant light on this difference. 

\subsection{Are Young Planets Larger than their Older Counterparts?}

\citet{Mann2016a} found that the $\simeq$800\,Myr old Hyades planet K2-25b has an unusually large radius when compared to other transiting planets from \kepler\ given its host star mass and incident flux. Higher-mass stars have larger disk masses \citep{Andrews:2013yq,Pascucci:2016aa} and hence are more likely to harbor large planets \citep[e.g.][]{Johnson:2010lr,Mulders2015}, and planets that receive more flux from their host star may lose their atmospheres faster (we use bolometric flux as a proxy for high-energy flux). Thus such large planetary radii suggest that, at this age, close-in planets around M-dwarfs may still be losing atmosphere due to interaction with their host star. 

\begin{figure*}[ht]
	\centering
	\includegraphics[width=0.95\textwidth]{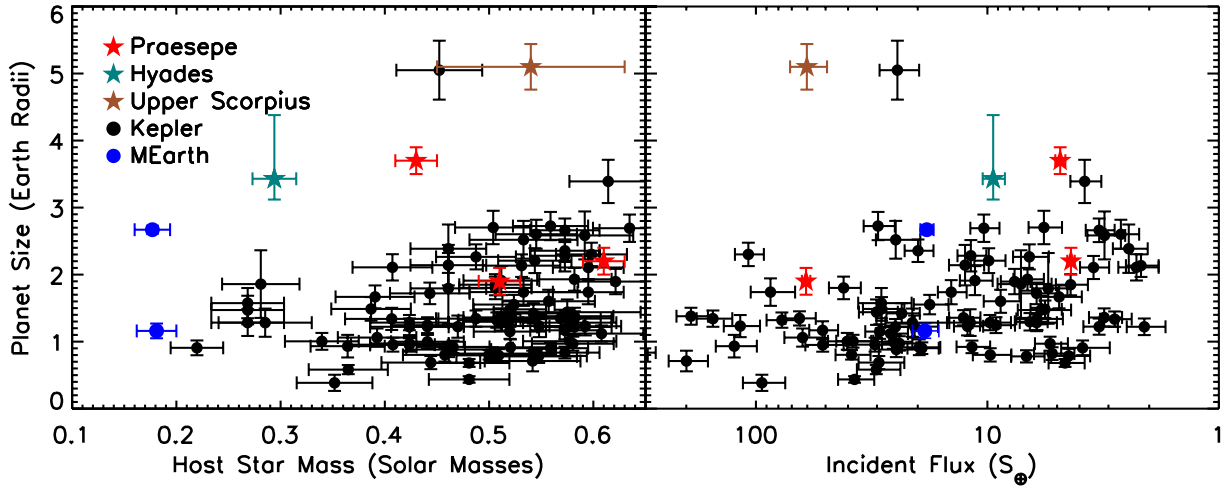} 
	\caption{Planet size versus stellar mass (left) and incident flux on the planet (right) for transiting planets orbiting $M_*<0.65M_\odot$ stars and with orbital periods $<30$\, days taken from the transit surveys MEarth, \ktwo\ and \kepler. Young systems are shown as stars while older ones are circles. K2-33b (Upper Scorpius) appears inflated, but is much younger ($\simeq$11\,Myr versus 800\,Myr) than systems considered here. Both K2-25b (Hyades) and K2-95b are atypically large given their host star mass. The other two Praesepe planets do not appear inflated, though for one (K2-104) this may be due to a higher incident flux. EPIC 211901114b is excluded, despite its large assigned radius ($9.6^{+5.3}_{-4.8}R_\earth$) because of its uncertain status and parameters.}
	\label{fig:praesepe_size}
\end{figure*}

We show a comparison of planet size as a function of host star mass and incident flux for M-dwarf planets drawn from transit surveys in Figure~\ref{fig:praesepe_size}. Stellar and planetary parameters for \kepler\ systems are taken from \citet{Gaidos2016b}, MEarth systems from \citet{2013A&A...551A..48A} and \citet{Berta-Thompson2015}, the Hyades system from \citet{Mann2016a}, and the Upper Scorpius system from \citet{Mann2016b}. Planets identified through RV surveys are not included becuase they have different selection criteria and observational biases, though GJ 436b and GJ 3470b are notable because they also have relatively large radii ($\simeq4R_\earth$) despite orbiting old ($>1$\,Gyr) stars \citep{von-Braun:2012lq,2014MNRAS.443.1810B}. GJ 436b, interestingly, also shows evidence of an evaporating atmosphere \citep{Ehrenreich2015}.

Excluding EPIC 211901114b, whose parameters are poorly constrained, only one target of six, K2-95b, is significantly larger than the $R_P-M_*$ sequence from \kepler, with a radius of 3.7$\pm$0.2R$_\earth$ and host star mass of $M_*=0.430\pm0.02M_\odot$, as noted by \citet{Obermeier:2016aa}. Although K2-95b looks like less of an outlier given its level of incident flux. The other planets around M-dwarfs are less remarkable in terms of their size. The more typical size for K2-104b may be due to a much higher level of flux from the host star stripping the atmosphere away faster than 800\,Myr. Close-in $\simeq1R_\earth$ planets, which are common in the \kepler\ M-dwarf sample \citep{Dressing2013,Dressing2015}, are likely missing from our sample due to detection limits; Praesepe targets are statistically more distant/faint than those observed by \kepler\ \citep{Gaidos2013}, and the shorter observing window yields fewer transits and S/N. It is still suggestive that 2-3 (depending on the status of EPIC~211901114b) of the 5 known planets orbiting $<$800\,Myr old M-dwarfs are large compared to 1-2 planets out of 90 orbiting old M-dwarfs in the \kepler-prime field. Mass determinations of these planets would be useful to determine if they have lower densities than their older counterparts, which would favor a scenario where close-in planets continue to lose atmosphere past 800\,Myr.

\acknowledgements
The authors thank Tim Morton for his help with the \textit{vespa} software. We also thank Marshall Johnson for his help with the transit fits. 

AWM was supported through Hubble Fellowship grant 51364 awarded by the Space Telescope Science Institute, which is operated by the Association of Universities for Research in Astronomy, Inc., for NASA, under contract NAS 5-26555. This research was supported by NASA grant NNX11AC33G to EG. A.V. is supported by the NSF Graduate Research Fellowship, Grant No. DGE 1144152. 

This work used the Immersion Grating Infrared Spectrograph (IGRINS) that was developed under a collaboration between the University of Texas at Austin and the Korea Astronomy and Space Science Institute (KASI) with the financial support of the US National Science Foundation under grant ASTR1229522, of the University of Texas at Austin, and of the Korean GMT Project of KASI. The IGRINS pipeline package PLP was developed by Dr. Jae-Joon Lee at Korea Astronomy and Space Science Institute and Professor Soojong Pak's team at Kyung Hee University. SNIFS on the UH 2.2-m telescope is part of the Nearby Supernova Factory project, a scientific collaboration among the Centre de Recherche Astronomique de Lyon, Institut de Physique Nucl\'{e}aire de Lyon, Laboratoire de Physique Nucl\'{e}aire et des Hautes Energies, Lawrence Berkeley National Laboratory, Yale University, University of Bonn, Max Planck Institute for Astrophysics, Tsinghua Center for Astrophysics, and the Centre de Physique des Particules de Marseille. Some of the data presented in this paper were obtained from the Mikulski Archive for Space Telescopes (MAST). STScI is operated by the Association of Universities for Research in Astronomy, Inc., under NASA contract NAS5-26555. Support for MAST for non-HST data is provided by the NASA Office of Space Science via grant NNX09AF08G and by other grants and contracts. This research was made possible through the use of the AAVSO Photometric All-Sky Survey (APASS), funded by the Robert Martin Ayers Sciences Fund. The authors acknowledge the Texas Advanced Computing Center (TACC) at The University of Texas at Austin for providing HPC resources that have contributed to the research results reported within this paper\footnote{\href{http://www.tacc.utexas.edu}{http://www.tacc.utexas.edu}}. These results made use of the Discovery Channel Telescope at Lowell Observatory. Lowell is a private, non-profit institution dedicated to astrophysical research and public appreciation of astronomy and operates the DCT in partnership with Boston University, the University of Maryland, the University of Toledo, Northern Arizona University and Yale University.

The authors wish to recognize and acknowledge the very significant cultural role and reverence that the summit of Maunakea has always had within the indigenous Hawaiian community. We are most fortunate to have the opportunity to conduct observations from this mountain.

\facilities{UH:2.2m (SNIFS), Keck:II (NIRC2), Smith (IGRINS), Kepler}

\bibliography{fullbiblio.bib}

\begin{thebibliography}{}
\expandafter\ifx\csname natexlab\endcsname\relax\def\natexlab#1{#1}\fi

\bibitem[{Adams \& Laughlin(2006)}]{Adams2006}
Adams, F.~C., \& Laughlin, G. 2006, The Astrophysical Journal, 649, 1004

\bibitem[{{Ag{\"u}eros} {et~al.}(2011){Ag{\"u}eros}, {Covey}, {Lemonias},
  {Law}, {Kraus}, {Batalha}, {Bloom}, {Cenko}, {Kasliwal}, {Kulkarni},
  {Nugent}, {Ofek}, {Poznanski}, \& {Quimby}}]{2011ApJ...740..110A}
{Ag{\"u}eros}, M.~A., {Covey}, K.~R., {Lemonias}, J.~J., {et~al.} 2011, \apj,
  740, 110

\bibitem[{{Ahn} {et~al.}(2012){Ahn}, {Alexandroff}, {Allende Prieto},
  {Anderson}, {Anderton}, {Andrews}, {Aubourg}, {Bailey}, {Balbinot}, {Barnes},
  \& et~al.}]{Ahn:2012kx}
{Ahn}, C.~P., {Alexandroff}, R., {Allende Prieto}, C., {et~al.} 2012, \apjs,
  203, 21

\bibitem[{{Aigrain} {et~al.}(2007){Aigrain}, {Hodgkin}, {Irwin}, {Hebb},
  {Irwin}, {Favata}, {Moraux}, \& {Pont}}]{2007MNRAS.375...29A}
{Aigrain}, S., {Hodgkin}, S., {Irwin}, J., {et~al.} 2007, \mnras, 375, 29

\bibitem[{{Aigrain} {et~al.}(2016){Aigrain}, {Parviainen}, \&
  {Pope}}]{2016MNRAS.459.2408A}
{Aigrain}, S., {Parviainen}, H., \& {Pope}, B.~J.~S. 2016, \mnras, 459, 2408

\bibitem[{{Aldering} {et~al.}(2002){Aldering}, {Adam}, {Antilogus}, {Astier},
  {Bacon}, {Bongard}, {Bonnaud}, {Copin}, {Hardin}, {Henault}, {Howell},
  {Lemonnier}, {Levy}, {Loken}, {Nugent}, {Pain}, {Pecontal}, {Pecontal},
  {Perlmutter}, {Quimby}, {Schahmaneche}, {Smadja}, \&
  {Wood-Vasey}}]{Aldering2002}
{Aldering}, G., {Adam}, G., {Antilogus}, P., {et~al.} 2002, in Society of
  Photo-Optical Instrumentation Engineers (SPIE) Conference Series, Vol. 4836,
  Survey and Other Telescope Technologies and Discoveries, ed. J.~A. {Tyson} \&
  S.~{Wolff}, 61--72

\bibitem[{{Allard} {et~al.}(2012){Allard}, {Homeier}, \&
  {Freytag}}]{2012RSPTA.370.2765A}
{Allard}, F., {Homeier}, D., \& {Freytag}, B. 2012, Royal Society of London
  Philosophical Transactions Series A, 370, 2765

\bibitem[{{Andrews} {et~al.}(2013){Andrews}, {Rosenfeld}, {Kraus}, \&
  {Wilner}}]{Andrews:2013yq}
{Andrews}, S.~M., {Rosenfeld}, K.~A., {Kraus}, A.~L., \& {Wilner}, D.~J. 2013,
  \apj, 771, 129

\bibitem[{{Anglada-Escud{\'e}} {et~al.}(2013){Anglada-Escud{\'e}},
  {Rojas-Ayala}, {Boss}, {Weinberger}, \& {Lloyd}}]{2013A&A...551A..48A}
{Anglada-Escud{\'e}}, G., {Rojas-Ayala}, B., {Boss}, A.~P., {Weinberger},
  A.~J., \& {Lloyd}, J.~P. 2013, \aap, 551, A48

\bibitem[{{Ballard} \& {Johnson}(2016)}]{Ballard2016}
{Ballard}, S., \& {Johnson}, J.~A. 2016, \apj, 816, 66

\bibitem[{{Baraffe} {et~al.}(2015){Baraffe}, {Homeier}, {Allard}, \&
  {Chabrier}}]{2015A&A...577A..42B}
{Baraffe}, I., {Homeier}, D., {Allard}, F., \& {Chabrier}, G. 2015, \aap, 577,
  A42

\bibitem[{{Barros} {et~al.}(2016){Barros}, {Demangeon}, \&
  {Deleuil}}]{Barros:2016aa}
{Barros}, S.~C.~C., {Demangeon}, O., \& {Deleuil}, M. 2016, \aap, 594, A100

\bibitem[{{Batalha}(2014)}]{2014PNAS..11112647B}
{Batalha}, N.~M. 2014, Proceedings of the National Academy of Science, 111,
  12647

\bibitem[{{Batalha} {et~al.}(2010){Batalha}, {Borucki}, {Koch}, {Bryson},
  {Haas}, {Brown}, {Caldwell}, {Hall}, {Gilliland}, {Latham}, {Meibom}, \&
  {Monet}}]{Batalha:2010fk}
{Batalha}, N.~M., {Borucki}, W.~J., {Koch}, D.~G., {et~al.} 2010, \apjl, 713,
  L109

\bibitem[{{Becker} {et~al.}(2015){Becker}, {Vanderburg}, {Adams}, {Rappaport},
  \& {Schwengeler}}]{Becker2015}
{Becker}, J.~C., {Vanderburg}, A., {Adams}, F.~C., {Rappaport}, S.~A., \&
  {Schwengeler}, H.~M. 2015, \apjl, 812, L18

\bibitem[{{Berta-Thompson} {et~al.}(2015){Berta-Thompson}, {Irwin},
  {Charbonneau}, {Newton}, {Dittmann}, {Astudillo-Defru}, {Bonfils}, {Gillon},
  {Jehin}, {Stark}, {Stalder}, {Bouchy}, {Delfosse}, {Forveille}, {Lovis},
  {Mayor}, {Neves}, {Pepe}, {Santos}, {Udry}, \&
  {W{\"u}nsche}}]{Berta-Thompson2015}
{Berta-Thompson}, Z.~K., {Irwin}, J., {Charbonneau}, D., {et~al.} 2015, \nat,
  527, 204

\bibitem[{{Biddle} {et~al.}(2014){Biddle}, {Pearson}, {Crossfield}, {Fulton},
  {Ciceri}, {Eastman}, {Barman}, {Mann}, {Henry}, {Howard}, {Williamson},
  {Sinukoff}, {Dragomir}, {Vican}, {Mancini}, {Southworth}, {Greenberg},
  {Turner}, {Thompson}, {Taylor}, {Levine}, \& {Webber}}]{2014MNRAS.443.1810B}
{Biddle}, L.~I., {Pearson}, K.~A., {Crossfield}, I.~J.~M., {et~al.} 2014,
  \mnras, 443, 1810

\bibitem[{{Boesgaard} {et~al.}(2013){Boesgaard}, {Roper}, \&
  {Lum}}]{Boesgaard2013}
{Boesgaard}, A.~M., {Roper}, B.~W., \& {Lum}, M.~G. 2013, \apj, 775, 58

\bibitem[{{Borucki} {et~al.}(2010){Borucki}, {Koch}, {Basri}, {Batalha},
  {Brown}, {Caldwell}, {Caldwell}, {Christensen-Dalsgaard}, {Cochran},
  {DeVore}, {Dunham}, {Dupree}, {Gautier}, {Geary}, {Gilliland}, {Gould},
  {Howell}, {Jenkins}, {Kondo}, {Latham}, {Marcy}, {Meibom}, {Kjeldsen},
  {Lissauer}, {Monet}, {Morrison}, {Sasselov}, {Tarter}, {Boss}, {Brownlee},
  {Owen}, {Buzasi}, {Charbonneau}, {Doyle}, {Fortney}, {Ford}, {Holman},
  {Seager}, {Steffen}, {Welsh}, {Rowe}, {Anderson}, {Buchhave}, {Ciardi},
  {Walkowicz}, {Sherry}, {Horch}, {Isaacson}, {Everett}, {Fischer}, {Torres},
  {Johnson}, {Endl}, {MacQueen}, {Bryson}, {Dotson}, {Haas}, {Kolodziejczak},
  {Van Cleve}, {Chandrasekaran}, {Twicken}, {Quintana}, {Clarke}, {Allen},
  {Li}, {Wu}, {Tenenbaum}, {Verner}, {Bruhweiler}, {Barnes}, \&
  {Prsa}}]{Borucki2010}
{Borucki}, W.~J., {Koch}, D., {Basri}, G., {et~al.} 2010, Science, 327, 977

\bibitem[{{Boyajian} {et~al.}(2012{\natexlab{a}}){Boyajian}, {McAlister}, {van
  Belle}, {Gies}, {ten Brummelaar}, {von Braun}, {Farrington}, {Goldfinger},
  {O'Brien}, {Parks}, {Richardson}, {Ridgway}, {Schaefer}, {Sturmann},
  {Sturmann}, {Touhami}, {Turner}, \& {White}}]{Boyajian:2012fk}
{Boyajian}, T.~S., {McAlister}, H.~A., {van Belle}, G., {et~al.}
  2012{\natexlab{a}}, \apj, 746, 101

\bibitem[{{Boyajian} {et~al.}(2012{\natexlab{b}}){Boyajian}, {von Braun}, {van
  Belle}, {McAlister}, {ten Brummelaar}, {Kane}, {Muirhead}, {Jones}, {White},
  {Schaefer}, {Ciardi}, {Henry}, {L{\'o}pez-Morales}, {Ridgway}, {Gies}, {Jao},
  {Rojas-Ayala}, {Parks}, {Sturmann}, {Sturmann}, {Turner}, {Farrington},
  {Goldfinger}, \& {Berger}}]{Boyajian2012}
{Boyajian}, T.~S., {von Braun}, K., {van Belle}, G., {et~al.}
  2012{\natexlab{b}}, \apj, 757, 112

\bibitem[{{Brandt} \& {Huang}(2015)}]{Brandt2015}
{Brandt}, T.~D., \& {Huang}, C.~X. 2015, \apj, 807, 24

\bibitem[{{Buchhave} {et~al.}(2012){Buchhave}, {Latham}, {Johansen},
  {Bizzarro}, {Torres}, {Rowe}, {Batalha}, {Borucki}, {Brugamyer}, {Caldwell},
  {Bryson}, {Ciardi}, {Cochran}, {Endl}, {Esquerdo}, {Ford}, {Geary},
  {Gilliland}, {Hansen}, {Isaacson}, {Laird}, {Lucas}, {Marcy}, {Morse},
  {Robertson}, {Shporer}, {Stefanik}, {Still}, \&
  {Quinn}}]{2012Natur.486..375B}
{Buchhave}, L.~A., {Latham}, D.~W., {Johansen}, A., {et~al.} 2012, \nat, 486,
  375

\bibitem[{{Cardelli} {et~al.}(1989){Cardelli}, {Clayton}, \&
  {Mathis}}]{1989ApJ...345..245C}
{Cardelli}, J.~A., {Clayton}, G.~C., \& {Mathis}, J.~S. 1989, \apj, 345, 245

\bibitem[{{Cargile} {et~al.}(2014){Cargile}, {James}, {Pepper}, {Kuhn},
  {Siverd}, \& {Stassun}}]{2014ApJ...782...29C}
{Cargile}, P.~A., {James}, D.~J., {Pepper}, J., {et~al.} 2014, \apj, 782, 29

\bibitem[{{Chatterjee} {et~al.}(2008){Chatterjee}, {Ford}, {Matsumura}, \&
  {Rasio}}]{2008ApJ...686..580C}
{Chatterjee}, S., {Ford}, E.~B., {Matsumura}, S., \& {Rasio}, F.~A. 2008, \apj,
  686, 580

\bibitem[{{Cochran} {et~al.}(2002){Cochran}, {Hatzes}, \&
  {Paulson}}]{Cochran2002}
{Cochran}, W.~D., {Hatzes}, A.~P., \& {Paulson}, D.~B. 2002, \aj, 124, 565

\bibitem[{{Cohen} {et~al.}(2003){Cohen}, {Wheaton}, \&
  {Megeath}}]{2003AJ....126.1090C}
{Cohen}, M., {Wheaton}, W.~A., \& {Megeath}, S.~T. 2003, \aj, 126, 1090

\bibitem[{{Covey} {et~al.}(2016){Covey}, {Ag{\"u}eros}, {Law}, {Liu}, {Ahmadi},
  {Laher}, {Levitan}, {Sesar}, \& {Surace}}]{2016ApJ...822...81C}
{Covey}, K.~R., {Ag{\"u}eros}, M.~A., {Law}, N.~M., {et~al.} 2016, \apj, 822,
  81

\bibitem[{{Crockett} {et~al.}(2012){Crockett}, {Mahmud}, {Prato},
  {Johns-Krull}, {Jaffe}, {Hartigan}, \& {Beichman}}]{Crockett2012}
{Crockett}, C.~J., {Mahmud}, N.~I., {Prato}, L., {et~al.} 2012, \apj, 761, 164

\bibitem[{{Crossfield} {et~al.}(2016){Crossfield}, {Ciardi}, {Petigura},
  {Sinukoff}, {Schlieder}, {Howard}, {Beichman}, {Isaacson}, {Dressing},
  {Christiansen}, {Fulton}, {L{\'e}pine}, {Weiss}, {Hirsch}, {Livingston},
  {Baranec}, {Law}, {Riddle}, {Ziegler}, {Howell}, {Horch}, {Everett}, {Teske},
  {Martinez}, {Obermeier}, {Benneke}, {Scott}, {Deacon}, {Aller}, {Hansen},
  {Mancini}, {Ciceri}, {Brahm}, {Jord{\'a}n}, {Knutson}, {Henning}, {Bonnefoy},
  {Liu}, {Crepp}, {Lothringer}, {Hinz}, {Bailey}, {Skemer}, \&
  {Defrere}}]{Crossfield:2016aa}
{Crossfield}, I.~J.~M., {Ciardi}, D.~R., {Petigura}, E.~A., {et~al.} 2016,
  \apjs, 226, 7

\bibitem[{{Cushing} {et~al.}(2004){Cushing}, {Vacca}, \&
  {Rayner}}]{Cushing2004}
{Cushing}, M.~C., {Vacca}, W.~D., \& {Rayner}, J.~T. 2004, \pasp, 116, 362

\bibitem[{{Dahm}(2015)}]{2015ApJ...813..108D}
{Dahm}, S.~E. 2015, \apj, 813, 108

\bibitem[{{Dahm} {et~al.}(2012){Dahm}, {Slesnick}, \& {White}}]{Dahm2012}
{Dahm}, S.~E., {Slesnick}, C.~L., \& {White}, R.~J. 2012, \apj, 745, 56

\bibitem[{{David} {et~al.}(2016{\natexlab{a}}){David}, {Hillenbrand},
  {Petigura}, {Carpenter}, {Crossfield}, {Hinkley}, {Ciardi}, {Howard},
  {Isaacson}, {Cody}, {Schlieder}, {Beichman}, \& {Barenfeld}}]{David2016b}
{David}, T.~J., {Hillenbrand}, L.~A., {Petigura}, E.~A., {et~al.}
  2016{\natexlab{a}}, \nat, 534, 658

\bibitem[{{David} {et~al.}(2016{\natexlab{b}}){David}, {Conroy}, {Hillenbrand},
  {Stassun}, {Stauffer}, {Rebull}, {Cody}, {Isaacson}, {Howard}, \&
  {Aigrain}}]{David2016}
{David}, T.~J., {Conroy}, K.~E., {Hillenbrand}, L.~A., {et~al.}
  2016{\natexlab{b}}, \aj, 151, 112

\bibitem[{{Delfosse} {et~al.}(2000){Delfosse}, {Forveille}, {S{\'e}gransan},
  {Beuzit}, {Udry}, {Perrier}, \& {Mayor}}]{Delfosse2000}
{Delfosse}, X., {Forveille}, T., {S{\'e}gransan}, D., {et~al.} 2000, \aap, 364,
  217

\bibitem[{{Dotter} {et~al.}(2008){Dotter}, {Chaboyer}, {Jevremovi{\'c}},
  {Kostov}, {Baron}, \& {Ferguson}}]{Dotter2008}
{Dotter}, A., {Chaboyer}, B., {Jevremovi{\'c}}, D., {et~al.} 2008, \apjs, 178,
  89

\bibitem[{{Douglas} {et~al.}(2016){Douglas}, {Ag{\"u}eros}, {Covey}, {Cargile},
  {Barclay}, {Cody}, {Howell}, \& {Kopytova}}]{2016ApJ...822...47D}
{Douglas}, S.~T., {Ag{\"u}eros}, M.~A., {Covey}, K.~R., {et~al.} 2016, \apj,
  822, 47

\bibitem[{{Douglas} {et~al.}(2014){Douglas}, {Ag{\"u}eros}, {Covey}, {Bowsher},
  {Bochanski}, {Cargile}, {Kraus}, {Law}, {Lemonias}, {Arce}, {Fierroz}, \&
  {Kundert}}]{2014ApJ...795..161D}
---. 2014, \apj, 795, 161

\bibitem[{{Dressing} \& {Charbonneau}(2013)}]{Dressing2013}
{Dressing}, C.~D., \& {Charbonneau}, D. 2013, \apj, 767, 95

\bibitem[{{Dressing} \& {Charbonneau}(2015)}]{Dressing2015}
---. 2015, \apj, 807, 45

\bibitem[{{Ehrenreich} {et~al.}(2015){Ehrenreich}, {Bourrier}, {Wheatley}, {Des
  Etangs}, {H{\'e}brard}, {Udry}, {Bonfils}, {Delfosse}, {D{\'e}sert}, {Sing},
  \& {Vidal-Madjar}}]{Ehrenreich2015}
{Ehrenreich}, D., {Bourrier}, V., {Wheatley}, P.~J., {et~al.} 2015, \nat, 522,
  459

\bibitem[{{Feiden} \& {Chaboyer}(2012)}]{Feiden2012a}
{Feiden}, G.~A., \& {Chaboyer}, B. 2012, \apj, 757, 42

\bibitem[{{Foreman-Mackey} {et~al.}(2013){Foreman-Mackey}, {Hogg}, {Lang}, \&
  {Goodman}}]{Foreman-Mackey2013}
{Foreman-Mackey}, D., {Hogg}, D.~W., {Lang}, D., \& {Goodman}, J. 2013, \pasp,
  125, 306

\bibitem[{{Fressin} {et~al.}(2013){Fressin}, {Torres}, {Charbonneau}, {Bryson},
  {Christiansen}, {Dressing}, {Jenkins}, {Walkowicz}, \&
  {Batalha}}]{Fressin:2013qy}
{Fressin}, F., {Torres}, G., {Charbonneau}, D., {et~al.} 2013, \apj, 766, 81

\bibitem[{{Gaidos} {et~al.}(2013){Gaidos}, {Fischer}, {Mann}, \&
  {Howard}}]{Gaidos2013b}
{Gaidos}, E., {Fischer}, D.~A., {Mann}, A.~W., \& {Howard}, A.~W. 2013, \apj,
  771, 18

\bibitem[{{Gaidos} \& {Mann}(2013)}]{Gaidos2013}
{Gaidos}, E., \& {Mann}, A.~W. 2013, \apj, 762, 41

\bibitem[{{Gaidos} {et~al.}(2016){Gaidos}, {Mann}, {Kraus}, \&
  {Ireland}}]{Gaidos2016b}
{Gaidos}, E., {Mann}, A.~W., {Kraus}, A.~L., \& {Ireland}, M. 2016, \mnras,
  457, 2877

\bibitem[{{Gaidos} {et~al.}(2014){Gaidos}, {Mann}, {L{\'e}pine}, {Buccino},
  {James}, {Ansdell}, {Petrucci}, {Mauas}, \& {Hilton}}]{Gaidos2014}
{Gaidos}, E., {Mann}, A.~W., {L{\'e}pine}, S., {et~al.} 2014, \mnras, 443, 2561

\bibitem[{{Gaidos} {et~al.}(2017){Gaidos}, {Mann}, {Rizzuto}, {Nofi}, {Mace},
  {Vanderburg}, {Feiden}, {Narita}, {Takeda}, {Esposito}, {De Rosa}, {Ansdell},
  {Hirano}, {Graham}, {Kraus}, \& {Jaffe}}]{Gaidos:2017aa}
{Gaidos}, E., {Mann}, A.~W., {Rizzuto}, A., {et~al.} 2017, \mnras, 464, 850

\bibitem[{{G{\'a}sp{\'a}r} {et~al.}(2009){G{\'a}sp{\'a}r}, {Rieke}, {Su},
  {Balog}, {Trilling}, {Muzzerole}, {Apai}, \& {Kelly}}]{2009ApJ...697.1578G}
{G{\'a}sp{\'a}r}, A., {Rieke}, G.~H., {Su}, K.~Y.~L., {et~al.} 2009, \apj, 697,
  1578

\bibitem[{{Gazak} {et~al.}(2012){Gazak}, {Johnson}, {Tonry}, {Dragomir},
  {Eastman}, {Mann}, \& {Agol}}]{Gazak:2012vn}
{Gazak}, J.~Z., {Johnson}, J.~A., {Tonry}, J., {et~al.} 2012, Advances in
  Astronomy, 2012, arXiv:1102.1036

\bibitem[{{Grunblatt} {et~al.}(2016){Grunblatt}, {Huber}, {Gaidos}, {Lopez},
  {Fulton}, {Vanderburg}, {Barclay}, {Fortney}, {Howard}, {Isaacson}, {Mann},
  {Petigura}, {Silva Aguirre}, \& {Sinukoff}}]{Grunblatt:2016aa}
{Grunblatt}, S.~K., {Huber}, D., {Gaidos}, E.~J., {et~al.} 2016, \aj, 152, 185

\bibitem[{{Hadden} \& {Lithwick}(2014)}]{2014ApJ...787...80H}
{Hadden}, S., \& {Lithwick}, Y. 2014, \apj, 787, 80

\bibitem[{{Hartkopf} {et~al.}(2001){Hartkopf}, {McAlister}, \&
  {Mason}}]{Hartkopf:2001}
{Hartkopf}, W.~I., {McAlister}, H.~A., \& {Mason}, B.~D. 2001, \aj, 122, 3480

\bibitem[{{Henden} {et~al.}(2012){Henden}, {Levine}, {Terrell}, {Smith}, \&
  {Welch}}]{Henden:2012fk}
{Henden}, A.~A., {Levine}, S.~E., {Terrell}, D., {Smith}, T.~C., \& {Welch}, D.
  2012, Journal of the American Association of Variable Star Observers
  (JAAVSO), 40, 430

\bibitem[{{Henry} \& {McCarthy}(1993)}]{Henry:1993fk}
{Henry}, T.~J., \& {McCarthy}, Jr., D.~W. 1993, \aj, 106, 773

\bibitem[{{H{\o}g} {et~al.}(2000){H{\o}g}, {Fabricius}, {Makarov}, {Urban},
  {Corbin}, {Wycoff}, {Bastian}, {Schwekendiek}, \& {Wicenec}}]{Hog2000}
{H{\o}g}, E., {Fabricius}, C., {Makarov}, V.~V., {et~al.} 2000, \aap, 355, L27

\bibitem[{{Howell} {et~al.}(2014){Howell}, {Sobeck}, {Haas}, {Still},
  {Barclay}, {Mullally}, {Troeltzsch}, {Aigrain}, {Bryson}, {Caldwell},
  {Chaplin}, {Cochran}, {Huber}, {Marcy}, {Miglio}, {Najita}, {Smith},
  {Twicken}, \& {Fortney}}]{Howell2014}
{Howell}, S.~B., {Sobeck}, C., {Haas}, M., {et~al.} 2014, \pasp, 126, 398

\bibitem[{{Huber} {et~al.}(2013){Huber}, {Carter}, {Barbieri}, {Miglio},
  {Deck}, {Fabrycky}, {Montet}, {Buchhave}, {Chaplin}, {Hekker},
  {Montalb{\'a}n}, {Sanchis-Ojeda}, {Basu}, {Bedding}, {Campante},
  {Christensen-Dalsgaard}, {Elsworth}, {Stello}, {Arentoft}, {Ford},
  {Gilliland}, {Handberg}, {Howard}, {Isaacson}, {Johnson}, {Karoff},
  {Kawaler}, {Kjeldsen}, {Latham}, {Lund}, {Lundkvist}, {Marcy}, {Metcalfe},
  {Silva Aguirre}, \& {Winn}}]{2013Sci...342..331H}
{Huber}, D., {Carter}, J.~A., {Barbieri}, M., {et~al.} 2013, Science, 342, 331

\bibitem[{{Husser} {et~al.}(2013){Husser}, {Wende-von Berg}, {Dreizler},
  {Homeier}, {Reiners}, {Barman}, \& {Hauschildt}}]{2013A&A...553A...6H}
{Husser}, T.-O., {Wende-von Berg}, S., {Dreizler}, S., {et~al.} 2013, \aap,
  553, A6

\bibitem[{{Jenkins} {et~al.}(2010){Jenkins}, {Caldwell}, {Chandrasekaran},
  {Twicken}, {Bryson}, {Quintana}, {Clarke}, {Li}, {Allen}, {Tenenbaum}, {Wu},
  {Klaus}, {Van Cleve}, {Dotson}, {Haas}, {Gilliland}, {Koch}, \&
  {Borucki}}]{Jenkins:2010qy}
{Jenkins}, J.~M., {Caldwell}, D.~A., {Chandrasekaran}, H., {et~al.} 2010,
  \apjl, 713, L120

\bibitem[{{Johnson} {et~al.}(2010){Johnson}, {Aller}, {Howard}, \&
  {Crepp}}]{Johnson:2010lr}
{Johnson}, J.~A., {Aller}, K.~M., {Howard}, A.~W., \& {Crepp}, J.~R. 2010,
  \pasp, 122, 905

\bibitem[{{Kennedy} \& {Kenyon}(2008)}]{2008ApJ...682.1264K}
{Kennedy}, G.~M., \& {Kenyon}, S.~J. 2008, \apj, 682, 1264

\bibitem[{{Kipping}(2010)}]{Kipping:2010lr}
{Kipping}, D.~M. 2010, \mnras, 408, 1758

\bibitem[{{Kipping}(2013)}]{Kipping2013}
---. 2013, \mnras, 435, 2152

\bibitem[{{Kipping}(2014)}]{2014MNRAS.444.2263K}
---. 2014, \mnras, 444, 2263

\bibitem[{{Kipping} {et~al.}(2012){Kipping}, {Dunn}, {Jasinski}, \&
  {Manthri}}]{2012MNRAS.421.1166K}
{Kipping}, D.~M., {Dunn}, W.~R., {Jasinski}, J.~M., \& {Manthri}, V.~P. 2012,
  \mnras, 421, 1166

\bibitem[{{Kov{\'a}cs} {et~al.}(2002){Kov{\'a}cs}, {Zucker}, \&
  {Mazeh}}]{Kovacs2002}
{Kov{\'a}cs}, G., {Zucker}, S., \& {Mazeh}, T. 2002, \aap, 391, 369

\bibitem[{{Kraus} \& {Hillenbrand}(2007)}]{Kraus:2007yq}
{Kraus}, A.~L., \& {Hillenbrand}, L.~A. 2007, \aj, 134, 2340

\bibitem[{{Kraus} {et~al.}(2016){Kraus}, {Ireland}, {Huber}, {Mann}, \&
  {Dupuy}}]{Kraus2016a}
{Kraus}, A.~L., {Ireland}, M.~J., {Huber}, D., {Mann}, A.~W., \& {Dupuy}, T.~J.
  2016, \aj, 152, 8

\bibitem[{{Kraus} {et~al.}(2008){Kraus}, {Ireland}, {Martinache}, \&
  {Lloyd}}]{Kraus2008}
{Kraus}, A.~L., {Ireland}, M.~J., {Martinache}, F., \& {Lloyd}, J.~P. 2008,
  \apj, 679, 762

\bibitem[{{Kreidberg}(2015)}]{Kreidberg2015}
{Kreidberg}, L. 2015, \pasp, 127, 1161

\bibitem[{{Lantz} {et~al.}(2004){Lantz}, {Aldering}, {Antilogus}, {Bonnaud},
  {Capoani}, {Castera}, {Copin}, {Dubet}, {Gangler}, {Henault}, {Lemonnier},
  {Pain}, {Pecontal}, {Pecontal}, \& {Smadja}}]{Lantz2004}
{Lantz}, B., {Aldering}, G., {Antilogus}, P., {et~al.} 2004, in Society of
  Photo-Optical Instrumentation Engineers (SPIE) Conference Series, Vol. 5249,
  Optical Design and Engineering, ed. L.~{Mazuray}, P.~J. {Rogers}, \&
  R.~{Wartmann}, 146--155

\bibitem[{Lee(2015)}]{IGRINS_plp}
Lee, J.-J. 2015, plp: Version 2.0, doi:10.5281/zenodo.18579

\bibitem[{{Libralato} {et~al.}(2016){Libralato}, {Nardiello}, {Bedin},
  {Borsato}, {Granata}, {Malavolta}, {Piotto}, {Ochner}, {Cunial}, \&
  {Nascimbeni}}]{2016MNRAS.tmp.1056L}
{Libralato}, M., {Nardiello}, D., {Bedin}, L.~R., {et~al.} 2016, \mnras,
  arXiv:1608.00459

\bibitem[{{Liu} {et~al.}(2016){Liu}, {Yong}, {Asplund}, {Ram{\'{\i}}rez}, \&
  {Mel{\'e}ndez}}]{Liu2016}
{Liu}, F., {Yong}, D., {Asplund}, M., {Ram{\'{\i}}rez}, I., \& {Mel{\'e}ndez},
  J. 2016, \mnras, 457, 3934

\bibitem[{{Lovis} \& {Mayor}(2007)}]{2007A&A...472..657L}
{Lovis}, C., \& {Mayor}, M. 2007, \aap, 472, 657

\bibitem[{{Luger} {et~al.}(2016){Luger}, {Agol}, {Kruse}, {Barnes}, {Becker},
  {Foreman-Mackey}, \& {Deming}}]{Luger:2016aa}
{Luger}, R., {Agol}, E., {Kruse}, E., {et~al.} 2016, \aj, 152, 100

\bibitem[{Mace {et~al.}(2016)Mace, Kim, Jaffe, Park, Lee, Kaplan, Yu, Yuk,
  Chun, Pak, Kim, Lee, Sneden, Afsar, Pavel, Lee, Oh, Jeong, Park, Kidder, Lee,
  Nguyen~Le, McLane, Gully-Santiago, Oh, Lee, Hwang, \& Park}]{Mace2016}
Mace, G., Kim, H., Jaffe, D.~T., {et~al.} 2016, 300 Nights of Science with
  IGRINS at McDonald Observatory, doi:10.1117/12.2232780

\bibitem[{{Malavolta} {et~al.}(2016){Malavolta}, {Nascimbeni}, {Piotto},
  {Quinn}, {Borsato}, {Granata}, {Bonomo}, {Marzari}, {Bedin}, {Rainer},
  {Desidera}, {Lanza}, {Poretti}, {Sozzetti}, {White}, {Latham}, {Cunial},
  {Libralato}, {Nardiello}, {Boccato}, {Claudi}, {Cosentino}, {Covino},
  {Gratton}, {Maggio}, {Micela}, {Molinari}, {Pagano}, {Smareglia}, {Affer},
  {Andreuzzi}, {Aparicio}, {Benatti}, {Bignamini}, {Borsa}, {Damasso}, {Di
  Fabrizio}, {Harutyunyan}, {Esposito}, {Fiorenzano}, {Gandolfi}, {Giacobbe},
  {Gonz{\'a}lez Hern{\'a}ndez}, {Maldonado}, {Masiero}, {Molinaro}, {Pedani},
  \& {Scandariato}}]{2016A&A...588A.118M}
{Malavolta}, L., {Nascimbeni}, V., {Piotto}, G., {et~al.} 2016, \aap, 588, A118

\bibitem[{{Malo} {et~al.}(2013){Malo}, {Doyon}, {Lafreni{\`e}re}, {Artigau},
  {Gagn{\'e}}, {Baron}, \& {Riedel}}]{Malo2013}
{Malo}, L., {Doyon}, R., {Lafreni{\`e}re}, D., {et~al.} 2013, \apj, 762, 88

\bibitem[{{Mandel} \& {Agol}(2002)}]{MandelAgol2002}
{Mandel}, K., \& {Agol}, E. 2002, \apjl, 580, L171

\bibitem[{{Mann} {et~al.}(2015){Mann}, {Feiden}, {Gaidos}, {Boyajian}, \& {von
  Braun}}]{Mann2015b}
{Mann}, A.~W., {Feiden}, G.~A., {Gaidos}, E., {Boyajian}, T., \& {von Braun},
  K. 2015, \apj, 804, 64

\bibitem[{{Mann} {et~al.}(2013{\natexlab{a}}){Mann}, {Gaidos}, \&
  {Ansdell}}]{Mann2013c}
{Mann}, A.~W., {Gaidos}, E., \& {Ansdell}, M. 2013{\natexlab{a}}, \apj, 779,
  188

\bibitem[{{Mann} {et~al.}(2013{\natexlab{b}}){Mann}, {Gaidos}, {Kraus}, \&
  {Hilton}}]{Mann2013b}
{Mann}, A.~W., {Gaidos}, E., {Kraus}, A., \& {Hilton}, E.~J.
  2013{\natexlab{b}}, \apj, 770, 43

\bibitem[{{Mann} \& {von Braun}(2015)}]{Mann2015a}
{Mann}, A.~W., \& {von Braun}, K. 2015, \pasp, 127, 102

\bibitem[{{Mann} {et~al.}(2016{\natexlab{a}}){Mann}, {Gaidos}, {Mace},
  {Johnson}, {Bowler}, {LaCourse}, {Jacobs}, {Vanderburg}, {Kraus}, {Kaplan},
  \& {Jaffe}}]{Mann2016a}
{Mann}, A.~W., {Gaidos}, E., {Mace}, G.~N., {et~al.} 2016{\natexlab{a}}, \apj,
  818, 46

\bibitem[{{Mann} {et~al.}(2016{\natexlab{b}}){Mann}, {Newton}, {Rizzuto},
  {Irwin}, {Feiden}, {Gaidos}, {Mace}, {Kraus}, {James}, {Ansdell},
  {Charbonneau}, {Covey}, {Ireland}, {Jaffe}, {Johnson}, {Kidder}, \&
  {Vanderburg}}]{Mann2016b}
{Mann}, A.~W., {Newton}, E.~R., {Rizzuto}, A.~C., {et~al.} 2016{\natexlab{b}},
  \aj, 152, 61

\bibitem[{{Markwardt}(2009)}]{Markwart2009}
{Markwardt}, C.~B. 2009, in Astronomical Society of the Pacific Conference
  Series, Vol. 411, Astronomical Data Analysis Software and Systems XVIII, ed.
  D.~A. {Bohlender}, D.~{Durand}, \& P.~{Dowler}, 251

\bibitem[{{McQuillan} {et~al.}(2013){McQuillan}, {Aigrain}, \&
  {Mazeh}}]{McQuillan:2013}
{McQuillan}, A., {Aigrain}, S., \& {Mazeh}, T. 2013, \mnras, 432, 1203

\bibitem[{{Meibom} {et~al.}(2013){Meibom}, {Torres}, {Fressin}, {Latham},
  {Rowe}, {Ciardi}, {Bryson}, {Rogers}, {Henze}, {Janes}, {Barnes}, {Marcy},
  {Isaacson}, {Fischer}, {Howell}, {Horch}, {Jenkins}, {Schuler}, \&
  {Crepp}}]{Meibom2013}
{Meibom}, S., {Torres}, G., {Fressin}, F., {et~al.} 2013, \nat, 499, 55

\bibitem[{{Mermilliod} {et~al.}(2009){Mermilliod}, {Mayor}, \&
  {Udry}}]{Mermilliod2009}
{Mermilliod}, J.-C., {Mayor}, M., \& {Udry}, S. 2009, \aap, 498, 949

\bibitem[{{Mochejska} {et~al.}(2002){Mochejska}, {Stanek}, {Sasselov}, \&
  {Szentgyorgyi}}]{2002AJ....123.3460M}
{Mochejska}, B.~J., {Stanek}, K.~Z., {Sasselov}, D.~D., \& {Szentgyorgyi},
  A.~H. 2002, \aj, 123, 3460

\bibitem[{{Morton}(2012)}]{2012ApJ...761....6M}
{Morton}, T.~D. 2012, \apj, 761, 6

\bibitem[{{Morton}(2015)}]{2015ascl.soft03011M}
---. 2015, {VESPA: False positive probabilities calculator}, Astrophysics
  Source Code Library, ascl:1503.011

\bibitem[{{Morton} \& {Winn}(2014)}]{Morton2014b}
{Morton}, T.~D., \& {Winn}, J.~N. 2014, \apj, 796, 47

\bibitem[{{Mulders} {et~al.}(2015){Mulders}, {Pascucci}, \&
  {Apai}}]{Mulders2015}
{Mulders}, G.~D., {Pascucci}, I., \& {Apai}, D. 2015, \apj, 798, 112

\bibitem[{{Mullally} {et~al.}(2015){Mullally}, {Coughlin}, {Thompson}, {Rowe},
  {Burke}, {Latham}, {Batalha}, {Bryson}, {Christiansen}, {Henze}, {Ofir},
  {Quarles}, {Shporer}, {Van Eylen}, {Van Laerhoven}, {Shah}, {Wolfgang},
  {Chaplin}, {Xie}, {Akeson}, {Argabright}, {Bachtell}, {Barclay}, {Borucki},
  {Caldwell}, {Campbell}, {Catanzarite}, {Cochran}, {Duren}, {Fleming},
  {Fraquelli}, {Girouard}, {Haas}, {He{\l}miniak}, {Howell}, {Huber}, {Larson},
  {Gautier}, {Jenkins}, {Li}, {Lissauer}, {McArthur}, {Miller}, {Morris},
  {Patil-Sabale}, {Plavchan}, {Putnam}, {Quintana}, {Ramirez}, {Silva Aguirre},
  {Seader}, {Smith}, {Steffen}, {Stewart}, {Stober}, {Still}, {Tenenbaum},
  {Troeltzsch}, {Twicken}, \& {Zamudio}}]{2015ApJS..217...31M}
{Mullally}, F., {Coughlin}, J.~L., {Thompson}, S.~E., {et~al.} 2015, \apjs,
  217, 31

\bibitem[{{Narita} {et~al.}(2010){Narita}, {Hirano}, {Sanchis-Ojeda}, {Winn},
  {Holman}, {Sato}, {Aoki}, \& {Tamura}}]{2010PASJ...62L..61N}
{Narita}, N., {Hirano}, T., {Sanchis-Ojeda}, R., {et~al.} 2010, \pasj, 62, L61

\bibitem[{{Netopil} {et~al.}(2016){Netopil}, {Paunzen}, {Heiter}, \&
  {Soubiran}}]{Netopil2016}
{Netopil}, M., {Paunzen}, E., {Heiter}, U., \& {Soubiran}, C. 2016, \aap, 585,
  A150

\bibitem[{{Neves} {et~al.}(2013){Neves}, {Bonfils}, {Santos}, {Delfosse},
  {Forveille}, {Allard}, \& {Udry}}]{Neves:2013yq}
{Neves}, V., {Bonfils}, X., {Santos}, N.~C., {et~al.} 2013, \aap, 551, A36

\bibitem[{{Nutzman} {et~al.}(2011){Nutzman}, {Fabrycky}, \&
  {Fortney}}]{2011ApJ...740L..10N}
{Nutzman}, P.~A., {Fabrycky}, D.~C., \& {Fortney}, J.~J. 2011, \apjl, 740, L10

\bibitem[{{Obermeier} {et~al.}(2016){Obermeier}, {Henning}, {Schlieder},
  {Crossfield}, {Petigura}, {Howard}, {Sinukoff}, {Isaacson}, {Ciardi},
  {David}, {Hillenbrand}, {Beichman}, {Howell}, {Horch}, {Everett}, {Hirsch},
  {Teske}, {Christiansen}, {L{\'e}pine}, {Aller}, {Liu}, {Saglia},
  {Livingston}, \& {Kluge}}]{Obermeier:2016aa}
{Obermeier}, C., {Henning}, T., {Schlieder}, J.~E., {et~al.} 2016, \aj, 152,
  223

\bibitem[{{Park} {et~al.}(2014){Park}, {Jaffe}, {Yuk}, {Chun}, {Pak}, {Kim},
  {Pavel}, {Lee}, {Oh}, {Jeong}, {Sim}, {Lee}, {Nguyen Le}, {Strubhar},
  {Gully-Santiago}, {Oh}, {Cha}, {Moon}, {Park}, {Brooks}, {Ko}, {Han}, {Nah},
  {Hill}, {Lee}, {Barnes}, {Yu}, {Kaplan}, {Mace}, {Kim}, {Lee}, {Hwang}, \&
  {Park}}]{Park2014}
{Park}, C., {Jaffe}, D.~T., {Yuk}, I.-S., {et~al.} 2014, in Society of
  Photo-Optical Instrumentation Engineers (SPIE) Conference Series, Vol. 9147,
  Society of Photo-Optical Instrumentation Engineers (SPIE) Conference Series,
  1

\bibitem[{{Parviainen} \& {Aigrain}(2015)}]{2015MNRAS.453.3821P}
{Parviainen}, H., \& {Aigrain}, S. 2015, \mnras, 453, 3821

\bibitem[{{Pascucci} {et~al.}(2016){Pascucci}, {Testi}, {Herczeg}, {Long},
  {Manara}, {Hendler}, {Mulders}, {Krijt}, {Ciesla}, {Henning}, {Mohanty},
  {Drabek-Maunder}, {Apai}, {Sz{\H u}cs}, {Sacco}, \&
  {Olofsson}}]{Pascucci:2016aa}
{Pascucci}, I., {Testi}, L., {Herczeg}, G.~J., {et~al.} 2016, \apj, 831, 125

\bibitem[{{Paulson} {et~al.}(2004){Paulson}, {Cochran}, \&
  {Hatzes}}]{2004AJ....127.3579P}
{Paulson}, D.~B., {Cochran}, W.~D., \& {Hatzes}, A.~P. 2004, \aj, 127, 3579

\bibitem[{{Perryman} {et~al.}(1998){Perryman}, {Brown}, {Lebreton}, {Gomez},
  {Turon}, {Cayrel de Strobel}, {Mermilliod}, {Robichon}, {Kovalevsky}, \&
  {Crifo}}]{1998A&A...331...81P}
{Perryman}, M.~A.~C., {Brown}, A.~G.~A., {Lebreton}, Y., {et~al.} 1998, \aap,
  331, 81

\bibitem[{{Petigura} {et~al.}(2013){Petigura}, {Marcy}, \&
  {Howard}}]{2013ApJ...770...69P}
{Petigura}, E.~A., {Marcy}, G.~W., \& {Howard}, A.~W. 2013, \apj, 770, 69

\bibitem[{{Pinsonneault} {et~al.}(2012){Pinsonneault}, {An},
  {Molenda-{\.Z}akowicz}, {Chaplin}, {Metcalfe}, \&
  {Bruntt}}]{Pinsonneault:2012lr}
{Pinsonneault}, M.~H., {An}, D., {Molenda-{\.Z}akowicz}, J., {et~al.} 2012,
  \apjs, 199, 30

\bibitem[{{Pont} {et~al.}(2006){Pont}, {Zucker}, \&
  {Queloz}}]{2006MNRAS.373..231P}
{Pont}, F., {Zucker}, S., \& {Queloz}, D. 2006, \mnras, 373, 231

\bibitem[{{Pope} {et~al.}(2016){Pope}, {Parviainen}, \&
  {Aigrain}}]{2016MNRAS.461.3399P}
{Pope}, B.~J.~S., {Parviainen}, H., \& {Aigrain}, S. 2016, \mnras, 461, 3399

\bibitem[{{Quinn} {et~al.}(2012){Quinn}, {White}, {Latham}, {Buchhave},
  {Cantrell}, {Dahm}, {F{\H u}r{\'e}sz}, {Szentgyorgyi}, {Geary}, {Torres},
  {Bieryla}, {Berlind}, {Calkins}, {Esquerdo}, \&
  {Stefanik}}]{2012ApJ...756L..33Q}
{Quinn}, S.~N., {White}, R.~J., {Latham}, D.~W., {et~al.} 2012, \apjl, 756, L33

\bibitem[{{Quinn} {et~al.}(2014){Quinn}, {White}, {Latham}, {Buchhave},
  {Torres}, {Stefanik}, {Berlind}, {Bieryla}, {Calkins}, {Esquerdo}, {F{\H
  u}r{\'e}sz}, {Geary}, \& {Szentgyorgyi}}]{Quinn2014}
---. 2014, \apj, 787, 27

\bibitem[{{Ram{\'{\i}}rez} \& {Mel{\'e}ndez}(2005)}]{2005ApJ...626..465R}
{Ram{\'{\i}}rez}, I., \& {Mel{\'e}ndez}, J. 2005, \apj, 626, 465

\bibitem[{{Raymond} {et~al.}(2009){Raymond}, {Barnes}, {Veras}, {Armitage},
  {Gorelick}, \& {Greenberg}}]{2009ApJ...696L..98R}
{Raymond}, S.~N., {Barnes}, R., {Veras}, D., {et~al.} 2009, \apjl, 696, L98

\bibitem[{{Reiners} {et~al.}(2010){Reiners}, {Bean}, {Huber}, {Dreizler},
  {Seifahrt}, \& {Czesla}}]{2010ApJ...710..432R}
{Reiners}, A., {Bean}, J.~L., {Huber}, K.~F., {et~al.} 2010, \apj, 710, 432

\bibitem[{{Rizzuto} {et~al.}(2011){Rizzuto}, {Ireland}, \&
  {Robertson}}]{2011MNRAS.416.3108R}
{Rizzuto}, A.~C., {Ireland}, M.~J., \& {Robertson}, J.~G. 2011, \mnras, 416,
  3108

\bibitem[{{Salaris} {et~al.}(2004){Salaris}, {Weiss}, \&
  {Percival}}]{2004A&A...414..163S}
{Salaris}, M., {Weiss}, A., \& {Percival}, S.~M. 2004, \aap, 414, 163

\bibitem[{{Scargle}(1981)}]{Scargle1981}
{Scargle}, J.~D. 1981, \apjs, 45, 1

\bibitem[{{Schneider} {et~al.}(2014){Schneider}, {Izzard}, {de Mink}, {Langer},
  {Stolte}, {de Koter}, {Gvaramadze}, {Hu{\ss}mann}, {Liermann}, \&
  {Sana}}]{2014ApJ...780..117S}
{Schneider}, F.~R.~N., {Izzard}, R.~G., {de Mink}, S.~E., {et~al.} 2014, \apj,
  780, 117

\bibitem[{{Seager} \& {Mall{\'e}n-Ornelas}(2003)}]{Seager:2003lr}
{Seager}, S., \& {Mall{\'e}n-Ornelas}, G. 2003, \apj, 585, 1038

\bibitem[{{Silva Aguirre} {et~al.}(2015){Silva Aguirre}, {Davies}, {Basu},
  {Christensen-Dalsgaard}, {Creevey}, {Metcalfe}, {Bedding}, {Casagrande},
  {Handberg}, {Lund}, {Nissen}, {Chaplin}, {Huber}, {Serenelli}, {Stello}, {Van
  Eylen}, {Campante}, {Elsworth}, {Gilliland}, {Hekker}, {Karoff}, {Kawaler},
  {Kjeldsen}, \& {Lundkvist}}]{2015MNRAS.452.2127S}
{Silva Aguirre}, V., {Davies}, G.~R., {Basu}, S., {et~al.} 2015, \mnras, 452,
  2127

\bibitem[{{Skrutskie} {et~al.}(2006){Skrutskie}, {Cutri}, {Stiening},
  {Weinberg}, {Schneider}, {Carpenter}, {Beichman}, {Capps}, {Chester},
  {Elias}, {Huchra}, {Liebert}, {Lonsdale}, {Monet}, {Price}, {Seitzer},
  {Jarrett}, {Kirkpatrick}, {Gizis}, {Howard}, {Evans}, {Fowler}, {Fullmer},
  {Hurt}, {Light}, {Kopan}, {Marsh}, {McCallon}, {Tam}, {Van Dyk}, \&
  {Wheelock}}]{Skrutskie2006}
{Skrutskie}, M.~F., {Cutri}, R.~M., {Stiening}, R., {et~al.} 2006, \aj, 131,
  1163

\bibitem[{{Stassun} {et~al.}(2014){Stassun}, {Pepper}, {Oelkers}, {Paegert},
  {De Lee}, \& {Sanchis-Ojeda}}]{2014arXiv1410.6379S}
{Stassun}, K.~G., {Pepper}, J.~A., {Oelkers}, R., {et~al.} 2014, ArXiv
  e-prints, arXiv:1410.6379

\bibitem[{{Taylor}(2006)}]{Taylor2006}
{Taylor}, B.~J. 2006, \aj, 132, 2453

\bibitem[{{Vacca} {et~al.}(2003){Vacca}, {Cushing}, \& {Rayner}}]{Vacca2003}
{Vacca}, W.~D., {Cushing}, M.~C., \& {Rayner}, J.~T. 2003, \pasp, 115, 389

\bibitem[{{Van Cleve} {et~al.}(2016){Van Cleve}, {Howell}, {Smith}, {Clarke},
  {Thompson}, {Bryson}, {Lund}, {Handberg}, \& {Chaplin}}]{2016PASP..128g5002V}
{Van Cleve}, J.~E., {Howell}, S.~B., {Smith}, J.~C., {et~al.} 2016, \pasp, 128,
  075002

\bibitem[{{Van Eylen} \& {Albrecht}(2015)}]{Van-Eylen2015}
{Van Eylen}, V., \& {Albrecht}, S. 2015, \apj, 808, 126

\bibitem[{{van Leeuwen}(2009)}]{van-Leeuwen2009}
{van Leeuwen}, F. 2009, \aap, 497, 209

\bibitem[{{van Saders} \& {Gaudi}(2011)}]{SandersGaudi2011}
{van Saders}, J.~L., \& {Gaudi}, B.~S. 2011, \apj, 729, 63

\bibitem[{{VandenBerg} \& {Clem}(2003)}]{2003AJ....126..778V}
{VandenBerg}, D.~A., \& {Clem}, J.~L. 2003, \aj, 126, 778

\bibitem[{{Vanderburg} \& {Johnson}(2014)}]{Vanderburg2014}
{Vanderburg}, A., \& {Johnson}, J.~A. 2014, \pasp, 126, 948

\bibitem[{{Vanderburg} {et~al.}(2016){Vanderburg}, {Latham}, {Buchhave},
  {Bieryla}, {Berlind}, {Calkins}, {Esquerdo}, {Welsh}, \&
  {Johnson}}]{2016ApJS..222...14V}
{Vanderburg}, A., {Latham}, D.~W., {Buchhave}, L.~A., {et~al.} 2016, \apjs,
  222, 14

\bibitem[{{von Braun} {et~al.}(2012){von Braun}, {Boyajian}, {Kane}, {Hebb},
  {van Belle}, {Farrington}, {Ciardi}, {Knutson}, {ten Brummelaar},
  {L{\'o}pez-Morales}, {McAlister}, {Schaefer}, {Ridgway}, {Collier Cameron},
  {Goldfinger}, {Turner}, {Sturmann}, \& {Sturmann}}]{von-Braun:2012lq}
{von Braun}, K., {Boyajian}, T.~S., {Kane}, S.~R., {et~al.} 2012, \apj, 753,
  171

\bibitem[{{Walkowicz} \& {Basri}(2013)}]{2013MNRAS.436.1883W}
{Walkowicz}, L.~M., \& {Basri}, G.~S. 2013, \mnras, 436, 1883

\bibitem[{{Wang} {et~al.}(2014){Wang}, {Fischer}, {Xie}, \&
  {Ciardi}}]{2014ApJ...791..111W}
{Wang}, J., {Fischer}, D.~A., {Xie}, J.-W., \& {Ciardi}, D.~R. 2014, \apj, 791,
  111

\bibitem[{{Wizinowich} {et~al.}(2000){Wizinowich}, {Acton}, {Shelton},
  {Stomski}, {Gathright}, {Ho}, {Lupton}, {Tsubota}, {Lai}, {Max}, {Brase},
  {An}, {Avicola}, {Olivier}, {Gavel}, {Macintosh}, {Ghez}, \&
  {Larkin}}]{2000PASP..112..315W}
{Wizinowich}, P., {Acton}, D.~S., {Shelton}, C., {et~al.} 2000, \pasp, 112, 315

\bibitem[{{Wright} {et~al.}(2010){Wright}, {Eisenhardt}, {Mainzer}, {Ressler},
  {Cutri}, {Jarrett}, {Kirkpatrick}, {Padgett}, {McMillan}, {Skrutskie},
  {Stanford}, {Cohen}, {Walker}, {Mather}, {Leisawitz}, {Gautier}, {McLean},
  {Benford}, {Lonsdale}, {Blain}, {Mendez}, {Irace}, {Duval}, {Liu}, {Royer},
  {Heinrichsen}, {Howard}, {Shannon}, {Kendall}, {Walsh}, {Larsen}, {Cardon},
  {Schick}, {Schwalm}, {Abid}, {Fabinsky}, {Naes}, \& {Tsai}}]{Wright:2010fk}
{Wright}, E.~L., {Eisenhardt}, P.~R.~M., {Mainzer}, A.~K., {et~al.} 2010, \aj,
  140, 1868

\bibitem[{{Yang} {et~al.}(2015){Yang}, {Chen}, \& {Zhao}}]{Yang2015}
{Yang}, X.~L., {Chen}, Y.~Q., \& {Zhao}, G. 2015, \aj, 150, 158

\bibitem[{{Yelda} {et~al.}(2010){Yelda}, {Lu}, {Ghez}, {Clarkson}, {Anderson},
  {Do}, \& {Matthews}}]{Yelda2010}
{Yelda}, S., {Lu}, J.~R., {Ghez}, A.~M., {et~al.} 2010, \apj, 725, 331

\bibitem[{{Yuk} {et~al.}(2010){Yuk}, {Jaffe}, {Barnes}, {Chun}, {Park}, {Lee},
  {Lee}, {Wang}, {Park}, {Pak}, {Strubhar}, {Deen}, {Oh}, {Seo}, {Pyo}, {Park},
  {Lacy}, {Goertz}, {Rand}, \& {Gully-Santiago}}]{2010SPIE.7735E..1MY}
{Yuk}, I.-S., {Jaffe}, D.~T., {Barnes}, S., {et~al.} 2010, in Society of
  Photo-Optical Instrumentation Engineers (SPIE) Conference Series, Vol. 7735,
  Society of Photo-Optical Instrumentation Engineers (SPIE) Conference Series

\bibitem[{{Zacharias} {et~al.}(2013){Zacharias}, {Finch}, {Girard}, {Henden},
  {Bartlett}, {Monet}, \& {Zacharias}}]{UCAC4}
{Zacharias}, N., {Finch}, C.~T., {Girard}, T.~M., {et~al.} 2013, \aj, 145, 44

\clearpage
\newpage
\end{thebibliography}

\global\pdfpageattr\expandafter{\the\pdfpageattr/Rotate 90}

\floattable
\rotate
\begin{deluxetable}{l l l l l l l l l l l }
\tablecaption{Planet/Transit-fit Parameters}
\tablewidth{0pt}
\tablehead{
\colhead{Parameter} & \colhead{K2-100b} & \colhead{K2-101b} & \colhead{K2-102b} & \colhead{K2-103b} & \colhead{K2-104b} & \colhead{EPIC 211901114\tablenotemark{a}} & \colhead{K2-95b} &
}
\startdata
 \multicolumn{8}{c}{Uniform stellar density Prior\tablenotemark{a}; $e$, $\omega$ fixed at 0} \\
\hline
Period (days) & $1.673915^{+0.000011}_{-0.000011}$ & $14.677303^{+0.000824}_{-0.000809}$ & $9.915651^{+0.001194}_{-0.001175}$ & $21.169687^{+0.001636}_{-0.001655}$ & $1.974189^{+0.000110}_{-0.000109}$ & $1.648932^{+0.000071}_{-0.000069}$ & $10.135097^{+0.000498}_{-0.000489}$ \\
$R_P/R_*$ & $0.0267^{+0.0011}_{-0.0005}$ & $0.0247^{+0.0012}_{-0.0007}$ & $0.0169^{+0.0010}_{-0.0008}$ & $0.0335^{+0.0015}_{-0.0011}$ & $0.0365^{+0.0029}_{-0.0015}$ & $0.1912^{+0.1053}_{-0.0948}$ & $0.0771^{+0.0031}_{-0.0021}$ \\
$T_0$\tablenotemark{b} (BJD-2400000) &$57144.06700^{+0.00027}_{-0.00027}$ & $57152.68125^{+0.00224}_{-0.00229}$ & $57139.65488^{+0.00538}_{-0.00539}$ & $57123.23787^{+0.00426}_{-0.00438}$ & $57140.38097^{+0.00252}_{-0.00257}$ & $57140.83259^{+0.00191}_{-0.00200}$ & $57140.74073^{+0.00204}_{-0.00211}$ \\
Density ($\rho_{\odot}$) &$2.24^{+0.34}_{-0.88}$ & $2.21^{+0.41}_{-0.89}$ & $2.65^{+0.72}_{-1.09}$ & $2.86^{+0.60}_{-1.11}$ & $3.48^{+1.26}_{-2.11}$ & $4.71^{+0.58}_{-0.58}$ & $3.59^{+0.70}_{-1.42}$ \\
Impact Parameter &$0.31^{+0.29}_{-0.22}$ & $0.32^{+0.29}_{-0.22}$ & $0.32^{+0.30}_{-0.22}$ & $0.31^{+0.28}_{-0.22}$ & $0.38^{+0.37}_{-0.26}$ & $1.11^{+0.11}_{-0.11}$ & $0.32^{+0.29}_{-0.22}$ \\
Duration (hr) &$1.61^{+0.02}_{-0.02}$ & $3.32^{+0.10}_{-0.09}$ & $2.68^{+0.14}_{-0.13}$ & $3.47^{+0.14}_{-0.12}$ & $1.44^{+0.13}_{-0.09}$ & $0.57^{+0.04}_{-0.03}$ & $2.64^{+0.11}_{-0.08}$ \\
$a/R_*$ &$7.8^{+0.4}_{-1.2}$  & $32.9^{+1.9}_{-5.2}$  & $26.9^{+2.2}_{-4.4}$  & $45.7^{+3.0}_{-6.9}$  & $10.0^{+1.1}_{-2.7}$  & $9.8^{+0.4}_{-0.4}$  & $30.2^{+1.8}_{-4.7}$  \\
Inclination (degrees) & $87.7^{+1.6}_{-2.9}$ & $89.5^{+0.4}_{-0.7}$ & $89.3^{+0.5}_{-0.9}$ & $89.6^{+0.3}_{-0.5}$ & $87.9^{+1.5}_{-3.6}$ & $83.5^{+0.7}_{-0.7}$ & $89.4^{+0.4}_{-0.8}$ \\
$R_P$\tablenotemark{c} ($R_\earth$) & $3.5^{+0.2}_{-0.2}$ & $2.0^{+0.1}_{-0.1}$ & $1.3^{+0.1}_{-0.1}$ & $2.2^{+0.2}_{-0.1}$ & $1.9^{+0.2}_{-0.1}$ & $9.6^{+5.3}_{-4.8}$ & $3.7^{+0.2}_{-0.2}$ \\
\hline
 \multicolumn{8}{c}{External stellar density prior; uniform priors on $\sqrt{e}\sin\omega$, $\sqrt{e}\cos \omega$} \\
\hline
Period (days) & $1.673916^{+0.000012}_{-0.000013}$ & $14.677286^{+0.000828}_{-0.000804}$ & $9.915615^{+0.001209}_{-0.001195}$ & $21.169619^{+0.001665}_{-0.001729}$ & $1.974190^{+0.000110}_{-0.000110}$ & \nodata & $10.135091^{+0.000495}_{-0.000488}$ \\
$R_P/R_*$ & $0.0269^{+0.0017}_{-0.0007}$ & $0.0247^{+0.0012}_{-0.0007}$ & $0.0170^{+0.0012}_{-0.0008}$ & $0.0336^{+0.0023}_{-0.0013}$ & $0.0365^{+0.0024}_{-0.0014}$ & \nodata & $0.0771^{+0.0033}_{-0.0020}$ \\
$T_0$\tablenotemark{b} (BJD-2400000) &$57144.06723^{+0.00055}_{-0.00037}$ & $57152.68135^{+0.00223}_{-0.00230}$ & $57139.65518^{+0.00552}_{-0.00552}$ & $57123.23803^{+0.00443}_{-0.00430}$ & $57140.38117^{+0.00268}_{-0.00261}$ & \nodata & $57140.74083^{+0.00205}_{-0.00208}$ \\
Density ($\rho_{\odot}$) &$0.71^{+0.10}_{-0.10}$ & $2.07^{+0.27}_{-0.28}$ & $2.15^{+0.27}_{-0.27}$ & $2.98^{+0.39}_{-0.40}$ & $4.58^{+0.63}_{-0.62}$ & \nodata & $5.03^{+0.73}_{-0.72}$ \\
Impact Parameter &$0.43^{+0.29}_{-0.30}$ & $0.35^{+0.25}_{-0.23}$ & $0.43^{+0.24}_{-0.27}$ & $0.38^{+0.33}_{-0.26}$ & $0.37^{+0.33}_{-0.25}$ & \nodata & $0.32^{+0.30}_{-0.22}$ \\
Duration (hr) &$1.55^{+0.23}_{-0.35}$ & $3.23^{+0.56}_{-0.63}$ & $2.66^{+0.56}_{-0.52}$ & $3.20^{+0.60}_{-1.02}$ & $1.27^{+0.21}_{-0.39}$ & \nodata & $2.28^{+0.31}_{-0.49}$ \\
$a/R_*$ &$6.2^{+0.3}_{-0.8}$  & $32.5^{+2.3}_{-2.4}$  & $25.3^{+1.9}_{-2.1}$  & $46.7^{+9.8}_{-4.5}$  & $11.0^{+2.4}_{-1.1}$  & \nodata & $33.2^{+6.8}_{-2.7}$  \\
Inclination (degrees) & $85.1^{+3.3}_{-3.1}$ & $89.4^{+0.4}_{-0.5}$ & $89.0^{+0.6}_{-0.6}$ & $89.5^{+0.3}_{-0.4}$ & $88.0^{+1.4}_{-1.9}$ & \nodata & $89.4^{+0.4}_{-0.5}$ \\
Eccentricity & $0.24^{+  0.19}_{-0.12}$  & $0.10^{+  0.18}_{-0.08}$  & $0.10^{+  0.16}_{-0.07}$  & $0.18^{+  0.27}_{-0.15}$  & $0.18^{+  0.29}_{-0.14}$  & \nodata & $0.16^{+  0.19}_{-0.11}$  \\
$\omega$ (degrees) &$29^{+ 41}_{-33}$  & $0^{+152}_{-118}$  & $-1^{+136}_{-132}$  & $0^{+156}_{-67}$  & $0^{+155}_{-63}$  & \nodata & $-2^{+157}_{-52}$  \\
$R_P$\tablenotemark{c} ($R_\earth$) & $3.5^{+0.2}_{-0.2}$ & $2.0^{+0.1}_{-0.1}$ & $1.3^{+0.1}_{-0.1}$ & $2.2^{+0.2}_{-0.1}$ & $1.9^{+0.2}_{-0.1}$ & \nodata & $3.7^{+0.2}_{-0.2}$ \\
\hline
\textit{vespa} FPP & 3.6x$10^{-3}$ & 1.4x$10^{-4}$ & 1.7x$10^{-3}$ & 1.9x$10^{-4}$ & 7.0x$10^{-3}$ & 0.02 & 1.3x$10^{-3}$ \\
\hline
\enddata
\tablenotetext{a}{For EPIC 211901114 the transit duration is unresolved, so we fix $e$ and $\omega$ to zero while simultaneously applying the Gaussian prior on $\rho$. Only one fit is done on this system.}
\tablenotetext{b}{BJD is given in Barycentric Dynamical Time (TBD) format.}
\tablenotetext{c}{Planet radius is derived using our stellar radius from Section~\ref{sec:params}.}
\tablecomments{Duration, $a/R_*$, and inclination are not fit as part of the MCMC; they are calculated from the fit parameters after the run is complete. }
\label{tab:planet}
\end{deluxetable}

\end{document}